\documentclass{JHEP3}
\usepackage{graphicx}
\usepackage{bbm}
\usepackage{amsmath}
\usepackage{amsfonts,amsbsy}
\usepackage{amssymb}
\usepackage{cite}

\def\myboldsymbol{\bf}

\newcommand{\nr}[1]{(\ref{#1})}
\newcommand{\fig}{fig.~}
\newcommand{\figs}{figs.~}
\newcommand{\eq}[1]{eq.~(#1)}
\newcommand{\eqpar}[1]{eq.~#1}
\newcommand{\beqa}{\begin{eqnarray}}
\newcommand{\eeqa}{\end{eqnarray}}
\def\qp{ {\bf q}_\perp } \def\pp{ {\bf p}_\perp }  \def\kp{ {\bf k}_\perp } \def\qs{Q_{\rm s}} \def\x{ {\bf x}
} \def\y{ {\bf y} } \def\z{ {\bf z} }  \def\p{ {\bf
    p} } \def\k{ {\bf k} } \def\q{ {\bf q} } \def\r{ {\bf r} }

\newcommand{\rt}{{\myboldsymbol{r}_\perp}}
\newcommand{\xt}{{\myboldsymbol{x}_\perp}}

\newcommand{\yt}{{\myboldsymbol{y}_\perp}}

\newcommand{\pt}{{\myboldsymbol{p}_\perp}}

\newcommand{\kt}{{\myboldsymbol{k}_\perp}} 
\newcommand{\ud}{\mathrm{d}}
\def\wt#1{\widetilde{#1}} \newcommand{\rb}[1]{ {\bf r}_{#1} }
\newcommand{\rts}[3]{ {\rho}_{#1}^{#2}(#3) }
\newcommand{\rtv}[2]{ {\rho}_{#1}^{#2} } \newcommand{\kpn}[1]{ {\bf
    k}_{#1\perp} } \newcommand{\sqrtv}[1]{ \sqrt{#1} } 

\title{Long range two-particle rapidity correlations in A+A collisions from high energy QCD evolution}

\author{Kevin Dusling\\Physics Department, Building 510A\\Brookhaven
  National Laboratory\\Upton, NY-11973, USA\\Email:
  \email{kdusling@quark.phy.bnl.gov}}

\author{Fran\c cois Gelis\\  Institut de Physique Th\'eorique (URA 2306 du CNRS)\\
  CEA/DSM/Saclay, B\^at. 774\\
  91191, Gif-sur-Yvette Cedex, France \\Email:
  \email{francois.gelis@cea.fr}}

\author{Tuomas Lappi\\ Department of Physics, P.O.~Box 35\\
  40014 University of Jyv\"askyl\"a, Finland\\ {\rm and} \\
Helsinki Institute of Physics, P.O.~Box 64,\\ 00014 University of Helsinki, Finland
\\Email:
  \email{tuomas.lappi@jyu.fi}}

\author{Raju Venugopalan\\Physics Department, Building
  510A\\Brookhaven National Laboratory\\Upton, NY-11973, USA\\Email:
  \email{rajuv@mac.com}}

\abstract{Long range rapidity correlations in A+A collisions are
  sensitive to strong color field dynamics at early times after the
  collision. These can be computed in a factorization
  formalism~\cite{GelisLV5} which expresses the $n$-gluon inclusive
  spectrum at arbitrary rapidity separations in terms of the
  multi-parton correlations in the nuclear wavefunctions. This
  formalism includes all radiative and rescattering contributions, to
  leading accuracy in $\alpha_s\Delta Y$, where $\Delta Y$ is the
  rapidity separation between either one of the measured gluons and a projectile,
  or between the measured gluons themselves.  In this paper, we use a mean field approximation for the
  evolution of the nuclear wavefunctions to obtain a compact result
  for inclusive two gluon correlations in terms of the unintegrated
  gluon distributions in the nuclear projectiles. The unintegrated
  gluon distributions satisfy the Balitsky-Kovchegov equation, which
  we solve with running coupling and with initial
  conditions constrained by existing data on electron-nucleus
  collisions. Our results are valid for arbitrary rapidity separations
  between measured gluons having transverse momenta
  $p_\perp,q_\perp\gtrsim \qs$, where $\qs$ is the saturation scale in the
  nuclear wavefunctions. We compare our results to data on long range
  rapidity correlations observed in the near-side ridge at RHIC and
  make predictions for similar long range rapidity correlations at the
  LHC.}
 
\begin{document}

\section{Introduction}

In a high energy heavy ion collision, several thousand particles are
produced in the initial interaction. The formation and evolution of
the resulting fireball can be described in a framework where the
incoming nuclei are sheets of strongly correlated coherent gluonic
fields called Color Glass Condensates
(CGC)~\cite{McLerV1,McLerV2,McLerV3,JalilKMW1,JalilKLW1,JalilKLW2,JalilKLW3,JalilKLW4,IancuLM1,IancuLM2,FerreILM1},
which are shattered in the collision to form strong classical fields
called the Glasma~\cite{LappiM1,GelisV4,GelisLV2}. The Glasma expands
and thermalizes to form a nearly perfect quark-gluon fluid, which
eventually hadronizes and freezes out to produce the large observed
multiplicity of particles.  While there is a fair amount of
circumstantial evidence on the temporal evolution of latter stages of
this space-time scenario, at present it is the earliest times, with
the strongest ``Glasma'' fields, that are most amenable to a
systematic theoretical treatment. This is because the early time
dynamics at times of order $1/\qs \lesssim 1$~fm is controlled by the
saturation scale $\qs$, which is the characteristic momentum scale in
the evolution of the bulk matter produced in the
collisions~\cite{GriboLR1,MuellQ1}.  Estimates for the magnitude of
$\qs^2$ are $1-1.4$~GeV$^2$ for gold nuclei at RHIC and $2.6-4$~GeV$^2$
for lead nuclei at LHC~\cite{KowalLV1}. The existence of this
semi-hard scale suggests that the Glasma may be described in weak
coupling, thereby opening a new window into the study of strongly
correlated quark-gluon matter.

The properties of the Glasma can be investigated by measuring long
range rapidity correlations of particles produced in the collision.
This is because the requirement that correlations be causal requires
the latest proper time $\tau_f$ that two particles could have been
correlated to be\footnote{This expression is valid in the scenario when the space-time
  rapidity and momentum space rapidities are strongly correlated.}
\begin{equation}
\tau_f = \tau_{\rm f.o.}\, \exp\left(-\frac{1}{2} \Delta y\right) \,,
\end{equation}
where the freezeout time $\tau_{\rm f.o.}$ is the proper time at which particles
from the fireball have no further interactions and $\Delta y$ is the
rapidity separation between the two particles. Thus, for 
$\tau_{\rm f.o.}\sim 10$~fm, two particles separated by 4
units in rapidity must have been correlated at no later than $1.4$~fm.
Strong correlations at space--time rapidity separations of $\Delta
\eta \leq 4$ units have been observed in the ``near-side ridge''
correlations measured by the PHOBOS experiment at RHIC~\cite{AlverA1}.
Correlations up to $\Delta \eta=1.5$ have been extensively studied by
the STAR collaboration~\cite{AbeleA1}. At the LHC, multi-particle
correlations at very large rapidity separations can be studied; these
are correlations that must have been created at {\it proper times well
  below a fermi}, thereby providing a unique window into the
non-linear dynamics of strong classical color fields in QCD.
 
It is therefore interesting to study the nature of these correlations
and what they reveal about the Glasma. Further, since these
correlations occur at very early times, after the collision, they are
closely related to multi-parton correlations in the nuclear
wavefunctions themselves. Multi-particle production in the Glasma, and
its relation to correlations in the nuclear wavefunctions can be
described in a weak coupling QCD framework where the degrees of
freedom are strong color sources $\rho^a \sim 1/g$ in the nuclei
(where $g$ is the QCD coupling constant) and gauge fields. Before the
collision, the distribution of sources and fields in the nuclear
wavefunctions evolves with rapidity; the evolution equations for the
color source distributions are the JIMWLK renormalization group
equations~\cite{JalilKMW1,JalilKLW1,JalilKLW2,JalilKLW3,JalilKLW4,IancuLM1,IancuLM2,FerreILM1}.
After the collision, the color sources become time dependent, thereby
enabling particle production in their radiation field. In
Refs.~\cite{GelisV2,GelisV3}, a field theory formalism was developed
to compute moments of the multiplicity distribution in the Glasma
systematically as an expansion in powers of $g^2$, while
simultaneously resumming contributions of order $g\rho \sim {\cal
  O}(1)$ from arbitrary numbers of insertions of color sources at each
order in $g^2$.

The naive expansion in powers of $g^2$ however breaks down because at
each order there are large logarithmic contributions in $x_{1,2}$, the
momentum fractions of partons in each of the nuclei, such that
$g^2\ln(1/x_{1,2})\sim {\cal O}(1)$ at small $x_{1,2}$. These
contributions therefore have to be resummed as well. In
Refs.~\cite{GelisLV3,GelisLV4,GelisLV5}, it was shown that inclusive
observables\footnote{This factorization is proven, to leading
  logarithmic accuracy in $x$, for inclusive multi-gluon spectra. It
  is straightforward to check that it applies to the expectation value
  of local operators such as the energy-momentum tensor $T^{\mu\nu}$,
  and multi-point correlations of such operators. It is unlikely to
  apply to exclusive final states that impose a veto on particle
  production in some regions of phase-space.} in the Glasma can be
expressed in a factorized form
\begin{eqnarray}
  \left<{\cal O}\right>_{_{\rm LLog}}
  =
  \int 
  \big[D\Omega_1(\bar{y},\xt)D\Omega_2(\bar{y},\xt)\big]\;
  W\big[\Omega_1(\bar{y},\xt)\big]W\big[\Omega_2(\bar{y},\xt)\big]\;
  {\cal O}_{_{\rm LO}}\; ,
\label{eq:fact-gen}
\end{eqnarray}
where the Wilson lines 
\begin{equation}
\label{eq:wilson-def}
\Omega_{1,2}(\bar{y},\x_\perp)\equiv
{\rm P}\,\exp ig \int_0^{x^\mp_y} \ud z^\mp\, 
\frac{1}{{\bf\nabla}_\perp^2}{\rho}_{1,2}(z^\mp,\x_\perp)\; .
\end{equation}
are ordered in rapidity (or, equivalently, in the longitudinal
coordinates $x^\mp$). In this definition of the Wilson line, the
rapidity $\bar{y}$ is measured from the fragmentation region of the
projectiles\footnote{It is therefore different from the usual laboratory frame 
rapidity $y$
used as a measure of the longitudinal momentum of a particle in the
final state of a reaction. For projectile $1$, moving in the
$+z$ direction, they are related by $\bar{y}=Y_{\rm beam}-y$,
and  for  projectile $2$ by $\bar{y}=-Y_{\rm beam}-y$.
The rapidity difference from the beam 
$\bar{y}$ is related to the upper bound $x^\mp$ in
\eq{\ref{eq:wilson-def}} by $\bar{y}=\ln(P^\pm x^\mp)$ where $P^\pm$ denotes 
the total longitudinal momentum of the projectiles 1 and 2 respectively.}. The
${\rho}_{1,2}$ are the color source densities of the nuclei in Lorenz
gauge at a given transverse co-ordinate $\xt$ and longitudinal
position $z^\mp$. The $W$'s are universal weight functionals 
(diagonal elements of density matrices) that give the probability distribution 
of a given configuration of sources (or equivalently the Wilson lines $\Omega_{1,2}$). 

If the separation scale between the sources and the fields in one of
the nuclei (moving in the + direction) is $\Lambda^+$, the requirement
that the physics be independent of this cutoff gives rise to the
JIMWLK renormalization group
equation~\cite{JalilKMW1,JalilKLW1,JalilKLW2,JalilKLW3,JalilKLW4,IancuLM1,IancuLM2,FerreILM1}
\begin{equation}
\frac{\partial}{\partial\ln(\Lambda^+)}W_{_{\Lambda^+}}[\Omega_1]
=
-{\cal H}_{\Lambda^+}
W_{_{\Lambda^+}}\big[\Omega_1\big]
\, ,
\label{eq:JIMWLK-evol}
\end{equation}
where $H_{\Lambda^+}$ is the JIMWLK Hamiltonian at the cutoff scale
$\Lambda^+$.  The precise form of the JIMWLK Hamiltonian is not needed
in the rest of this paper, and we refer the reader to
Refs.~\cite{JalilKMW1,JalilKLW1,JalilKLW2,JalilKLW3,JalilKLW4,IancuLM1,IancuLM2,FerreILM1,GelisLV3}
for further details. The distributions $W[\Omega_{1,2}]$ that enter in
\eq{\ref{eq:fact-gen}} are the limits when $\Lambda_\pm\to 0$ of the
cutoff dependent distributions $W_{\Lambda_\pm}[\Omega_{1,2}]$

The ``master'' formula in \eq{\ref{eq:fact-gen}} is valid for all
moments of the multiplicity distribution of gluons produced in the
Glasma. Given a non-perturbative initial condition, the $W$ weight
functionals encode information about gluon correlations at all
transverse positions and rapidities. A remarkable feature of
\eq{\ref{eq:fact-gen}} is that the only ``process dependent'' input on
the right hand side of the expression is the observable computed to
leading order\footnote{By leading order, we mean the first term in the
  expansion
  \begin{equation}
{\cal O}\left[\rho_1,\rho_2\right]
=
\frac{1}{g^{2n}}\Big[c_0+c_1 g^2+c_2 g^4+\cdots\Big]\; ,
\label{eq:expansion}
  \end{equation}
  where each term corresponds to a different loop order. Each of the
  coefficients $c_n$ is itself an infinite series of terms involving
  arbitrary orders in $(g\rho_{1,2})^p$. The ``leading order''
  contribution,
  \begin{equation}
{\cal O}_{_{\rm LO}}[\rho_1,\rho_2]\equiv\frac{c_0}{g^{2n}}\; ,
  \end{equation}
  corresponds to an infinite sum of tree diagrams.} in $\alpha_s$;
${\cal O}_{_{\rm LO}}$, albeit non-perturbative, is obtained by solving
classical Yang--Mills equations for the two nuclei and has been
studied extensively for the single
inclusive~\cite{KrasnV1,KrasnV2,KrasnV3,KrasnNV1,KrasnNV2,KrasnNV3,Lappi1,Lappi2,Lappi3}
gluon spectra using numerical lattice
methods. 

The Glasma Flux Tube picture of A+A collisions~\cite{DumitGMV1} is a
consequence of this master formula.  At short times after the
collision, the solutions of the Yang--Mills equations provide
longitudinal chromo-electric and chromo-magnetic fields in the forward
light cone.  In the McLerran--Venugopalan (MV)
model~\cite{McLerV1,McLerV2,McLerV3}, the distribution of color
correlations in the nuclear weight functionals $W$ is Gaussian with
its width set by the saturation scale $\qs$. The averaging over the
weight functionals in \eq{\ref{eq:fact-gen}} ensures that
only chromo-electric and magnetic fields localized in transverse areas
of size $1/\qs^2$ contribute to multi-particle production
obtained by averaging over all events.  In the MV model, color
correlations extend to arbitrarily large rapidity intervals; this
results in a picture of multi-particle production as arising from
boost invariant flux tubes of size $1/\qs^2$.  This picture, albeit
simple, gives a qualitative~\cite{DumitGMV1} and even
semi-quantitative~\cite{GavinMM1,MoschGM1} description of the
near-side ridge correlations observed in the RHIC heavy ion
experiments.

Reality however is more complex and the boost invariance of the Glasma
flux tubes is violated both by quantum evolution effects (real gluon
emissions and virtual corrections) in rapidity between the beam
rapidity and the rapidities of the measured gluons and likewise by {\it quantum
evolution between the measured gluons}.  When the maximum rapidity
interval between measured gluons $\Delta Y \ll 1/\alpha_s$, quantum
radiation between the observed gluons is not significant and correlated
gluon emission is approximately independent of $\Delta Y$. The
factorization formalism for this case was developed in
Refs.~\cite{GelisLV3,GelisLV4}. A quantitative understanding of how
correlations depend on $\Delta Y$, for arbitrary $\Delta Y \leq 2
Y_{\rm beam}$, requires that one understands the dynamics of real and
virtual quantum corrections between the tagged gluons. As demonstrated
in Ref.~\cite{GelisLV5}, all the necessary information is contained in
\eq{\ref{eq:fact-gen}} and the general formula for two gluon
correlations was derived in that paper.

In this paper, we will exploit the formalism of Ref.~\cite{GelisLV5}
to evaluate the rapidity dependence of correlated two gluon emission
in A+A collisions at RHIC and LHC energies. Because solving the JIMWLK
equation is highly computationally intensive~\cite{RummuW1}, we will
instead use a mean field approximation to this evolution equation
known as the Balitsky--Kovchegov (BK) equation~\cite{Balit2,Kovch2}.
The BK equation is a very good approximation to the JIMWLK
equation~\cite{KovchKRW1}, albeit it must be noted that this may be
true only for a limited class of observables. Recently, significant
progress has been made in computing Next-to-Leading-Order (NLO)
contributions to the BK equation~\cite{BalitC1,KovchW1} and the
results have been successfully applied to phenomenological studies of
HERA DIS data at small $x$~\cite{AlbacAMS1,AlbacAMS2}. To apply this
framework to nuclear collisions, we will first fix the initial
conditions for the running coupling BK evolution of unintegrated gluon
distributions for nuclei by fitting the existing inclusive e+A fixed
target data. We will then apply our results to compute the rapidity
dependence of two gluon correlations in A+A collisions.

Our paper is organized as follows. Readers interested primarily in the
results of the paper can proceed directly to section~\ref{sec:results}.
In section~\ref{sec:doubleinc},
we will restate the results of Ref.~\cite{GelisLV5} for the single and
double inclusive gluon spectrum and demonstrate how the expressions
simplify vastly in the mean field BK approximation of quantum
evolution. In section~\ref{sec:ktfact}, the correlated two gluon distribution is
expressed in terms of the unintegrated gluon distributions of nuclei.
These distributions are determined in section~\ref{sec:BK} from numerical
solutions of the  BK equation with running coupling.
To constrain the initial conditions
for the evolution of these nuclear unintegrated distributions, we fit
available fixed target e+A data --- the initial conditions for proton
unintegrated distributions were determined previously in global fits to the
HERA e+p data~\cite{AlbacAMS1,AlbacAMS2}. In section~\ref{sec:results}, we discuss our
results for correlated two gluon production in the context of RHIC
data from STAR and PHOBOS. We make predictions for what one may expect
at the LHC. The final section contains our conclusions.  Details of
the computations, solutions of running coupling BK equations and fits
to e+A fixed target data are contained in two appendices.

\section{Double inclusive gluon spectra at arbitrary rapidities}
\label{sec:doubleinc}

\begin{figure}[htbp]
\begin{center}
\includegraphics[scale=1]{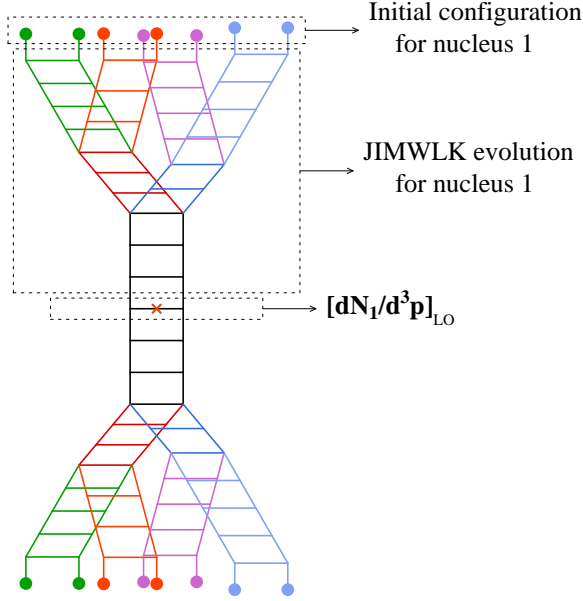}
\end{center}
\caption{\label{fig:1-gluon} Diagrammatic representation of the various building blocks in the
factorized formula for the inclusive single gluon spectrum. The lower part of
the figure, representing nucleus 2, is made up of identical building blocks.}
\end{figure}

The general leading log factorization formula~(\ref{eq:fact-gen})
gives the distribution of gluons at \emph{all} rapidities in the
problem~\cite{GelisLV5,Lappi5,GelisLV6}, including all the rapidity
correlations in the leading log approximation. For single and double
inclusive gluon production we only need the distribution of Wilson
lines at one or two gluon rapidities respectively. We shall now
specialize the generic formula~(\ref{eq:fact-gen}) to the this
specific case.  The single inclusive gluon spectrum $\ud N_1/\ud^3\p$
at leading order depends only on the Wilson lines\footnote{From here
  onwards, we shall use the standard definition of the rapidity $y$
  instead of the rapidity distance $\bar{y}$ from the projectile.
In terms of the Wilson lines introduced in the previous section, this translates to 
  $\Omega_1(y,\xt)=\Omega_1(Y_{\rm beam}-\bar{y},\xt)$ and
  $\Omega_2(y,\xt)=\Omega_2(\bar{y}+Y_{\rm beam},\xt)$.}
$\Omega_{1,2}(y,\xt)$ at the rapidity $y=y_p$ of the produced gluon
and not on the whole rapidity range contained in
\eq{\ref{eq:fact-gen}}.  Therefore, we can simplify
\eq{\ref{eq:fact-gen}} by inserting the identity
 \begin{equation}
 1=\int \big[DU_{1,2}(\xt)\big]\;
 \delta\big[U_{1,2}(\xt)-\Omega_{1,2}(y_p,\xt)\big]
\label{eq:identity}
 \end{equation} 
 and by defining the corresponding probability distributions for
 configurations of Wilson lines at the rapidity $y_p$
\begin{equation}
Z_{y_p}[U_{1,2}(\xt)]
\equiv
\int\big[D\Omega_{1,2}(y,\xt)\big]\;
W\big[\Omega_{1,2}(y,\xt)\big]
\;
\delta\big[U_{1,2}(\xt)-\Omega_{1,2}(y_p,\xt)\big]\; .
\label{eq:Z}
\end{equation}
One then obtains the all order leading log result for the single
inclusive gluon spectrum at the rapidity $y_p$ to be
\begin{equation}
\Big<\frac{\ud N_1}{\ud^2\p_\perp \ud y_p}\Big>_{_{\rm LLog}}
=
\int
\big[D U_{1}(\xt)\,DU_{2}(\xt)\big] \;
Z_{y_p}\left[U_{1}\right]\,
Z_{y_p}\left[U_{2}\right]\;
\left.
\frac{\ud N_1\big[U_{1},U_{2}\big]}{\ud^2\p_\perp \ud y_p}
\right|_{_{\rm LO}}\!.
\label{eq:N1-resummed}
\end{equation}
Note that the distribution $Z_{y_p}[U]$ obeys the JIMWLK
equation,
\begin{equation}
\partial_{y_p} Z_{y_p}[U] = {\cal H}_{y_p}\,Z_{y_p}[U]\; ,
\end{equation}
which must be supplemented by an initial condition at a rapidity close
to the fragmentation region of the projectiles.
Eq.~\ref{eq:N1-resummed} is illustrated in figure
\ref{fig:1-gluon}.

As will become clear shortly, it is more convenient to express
\eq{\ref{eq:N1-resummed}} in terms of color charge densities as 
\begin{equation}
\left< \frac{\ud N_1}{\ud^2\pp \ud y_p }\right>_{_{\rm LLog}}
=\int \left[D\rts{1}{p}{\xt} D\rts{2}{p}{\xt}
\right]Z_{y_p}\left[\rtv{1}{p}\right]
Z_{y_p}\left[\rtv{2}{p}\right]\;\left.
\frac{\ud N_1\left[\rtv{1}{p},\rtv{2}{p}\right]}{\ud^2\pp
  \ud y_p}\right|_{_{\rm LO}}\; .
\label{eq:single-inclusive-1}
\end{equation}
Here, the superscript on $\rho_{1,2}^p$ denotes the color charge
distribution of nucleus 1 or 2 evaluated at the rapidity $y_p$.

Following a similar though slightly more involved
derivation~\cite{GelisLV5}, one obtains from \eq{\ref{eq:fact-gen}}
the expression
\begin{eqnarray}
\left< \frac{\ud N_2}{\ud^2\pp \ud y_p \ud^2\qp \ud y_q}\right>_{_{\rm LLog}}
&=&
\int \left[D\rts{1}{p}{\xt} D\rts{2}{p}{\xt} 
D\rts{1}{q}{\xt} D\rts{2}{q}{\xt}\right]
\nonumber\\
&&\times
Z_{y_p}[\rtv{1}{p}]\,G_{y_p,y_q}[\rtv{1}{p},\rtv{1}{q}]\,
Z_{y_q}[\rtv{2}{q}]\,G_{y_q,y_p}[\rtv{2}{q},\rtv{2}{p}]
\nonumber\\
&&\times 
\left.
\frac{\ud N_1\left[\rtv{1}{p},\rtv{2}{p}\right]}{\ud^2\pp \ud y_p}
\right|_{_{\rm LO}}\;
\left.
\frac{\ud N_1\left[\rtv{1}{q},\rtv{2}{q}\right]}{\ud^2\qp \ud y_q}
\right|_{_{\rm LO}}\; .
\label{eq:double-inclusive-1}
\end{eqnarray}
It is important to note that here we have taken $y_q > y_p$, where $y_p$ is at an earlier
stage in the evolution of projectile $1$ and likewise, $y_q$ is at an earlier stage in the evolution
of projectile $2$. This convention will be
followed for the rest of this paper.
Also, in \eq{\ref{eq:double-inclusive-1}}, $G_{y_q,y_p}\big[\rho_{1,2}^q,\rho_{1,2}^p\big] $ is a Green's
function of the operator $\partial_y-{\cal H}_y$,
\begin{equation}
\partial_{y_q} G_{y_q,y_p}\big[U^q,U^p\big] 
= \mathcal{H}_{y_q}\,G_{y_q,y_p}\big[U^q,U^p\big]\; ,
\label{eq:green-JIMWLK}
\end{equation}
with the boundary condition
\begin{equation} 
\label{eq:greenic}
\lim_{y_q \to y_p} G_{y_q,y_p}\big[\rho^q,\rho^p\big] 
=
\delta\big[\rho^q-\rho^p\big]\; .
\end{equation}
This Green's function describes quantum evolution between two
specified rapidities, in the presence of strong color fields from the
projectiles. It relates the distribution of color sources at a given
rapidity with the distribution at another rapidity through the
relation
\begin{eqnarray}
\label{eq:green-Z}
Z_{y_q}\left[\rtv{1}{q}\right]
&=&
\int \left[D\rtv{1}{p}\right]\;
 Z_{y_p}\left[\rtv{1}{p}\right]\,G_{y_p,y_q}\left[\rtv{1}{p},\rtv{1}{q}\right]
\nonumber\\
Z_{y_p}\left[\rtv{2}{p}\right]
&=&
\int \left[D\rtv{2}{q}\right]\; 
Z_{y_q}\left[\rtv{2}{q}\right]\,G_{y_q,y_p}\left[\rtv{2}{q},\rtv{2}{p}\right]
\; .
\end{eqnarray}

At this stage, it is important to note that the BK mean
field form of the Balitsky-JIMWLK evolution equation for the two point
Wilson line correlator (the ``dipole cross-section'')
requires that the correlator of the product of traces of two pairs of
Wilson lines factorizes into the product of the correlators of traces
of pairs of Wilson lines\footnote{These Wilson lines are defined in
  terms of the color charge densities through
  \eq{\ref{eq:wilson-def}} projected on to a particular rapidity
  $\wt{U}(\x_\perp)\equiv \exp\left( ig
    \frac{1}{\boldsymbol\nabla_\perp^2}{\rho}_{a}(\x_\perp)t^a\right)$,
  where the $t^a$'s are the generators of the fundamental
  representation of $SU(N_c)$.  The corresponding expression in the
  adjoint representation is given in \eq{\ref{eq:pathGlue}}.},
\begin{eqnarray}
&& \left< \,{\rm tr}\,[{\wt U}^\dagger({\x_\perp}) {\wt U}({\z_\perp})]
\,\,{\rm tr}\,[{\wt U}^\dagger({\z_\perp}) {\wt U}({\y_\perp})]\,\right>
\nonumber \\
&&\qquad\qquad=
\left< \,{\rm tr}\,[{\wt U}^\dagger({\x_\perp}) {\wt U}({\z_\perp})]\,\right>
\,
\left<\,{\rm tr}\,[{\wt U}^\dagger({\z_\perp}) {\wt U}({\y_\perp})]\,\right> 
\; ,
\label{eq:BK-Factorization}
\end{eqnarray}
to leading order in a $1/N_c$ expansion. The averages
$\big<\cdots\big>$ are performed over the color sources of a large nucleus
with the weight functional $Z_y$. The factorization in
\eq{\ref{eq:BK-Factorization}} can be achieved with Gaussian
correlations among the color sources. Note however that we need a
non-local Gaussian distribution to accommodate the quantum BK
evolution~\cite{FujiiGV2}. One obtains therefore\footnote{It is to be
  understood that repeated color indices $a$ are summed over.}
\begin{eqnarray}
Z_{y_p}\left[\rtv{1,2}{p}\right]
=
\exp\left[-\frac{1}{2}\int_{\xt,\yt}
\frac{\rts{1,2}{a,p}{\xt}\rts{1,2}{a,p}{\yt}}
{\mu_{A_{1,2}}^2(y_p,\xt-\yt)}\right]\,,
\label{eq:Z1}
\end{eqnarray}
where $\mu_{A_{1,2}}^2(y_p,x_\perp-y_\perp)$ represents the color
charge squared per unit area of nucleus 1 or nucleus 2 as seen by a
particle having rapidity $y_p$.  Even though in this work we will consider collisions between identical nuclei, we shall retain the
explicit $A_{1,2}$ notation for generality.

From \eq{\ref{eq:green-Z}}, because the $Z_y$ functionals on the l.h.s
and the r.h.s are both Gaussians, the Green's function $G_{y_q,y_p}$
must be Gaussian as well.  One obtains
\begin{eqnarray}
\label{eq:Green-Gauss}
G_{y_p,y_q}\left[\rtv{1}{p},\rtv{1}{q}\right]
&=&
\exp\left[-\frac{1}{2}\int_{\xt,\yt}
\frac{\Delta\rts{1}{}{\xt} \Delta\rts{1}{}{\yt}}
{\Delta\mu_{A_1}^2(\xt-\yt)}\right]\nonumber\\
G_{y_q,y_p}\left[\rtv{2}{q},\rtv{2}{p}\right]
&=&
\exp\left[-\frac{1}{2}\int_{\xt,\yt}
\frac{\Delta\rts{2}{}{\xt} \Delta\rts{2}{}{\yt}}
{\Delta\mu_{A_2}^2(\xt-\yt)}\right]\; ,
\label{eq:gaussian-G}
\end{eqnarray}
where we have defined 
\begin{eqnarray}
\label{eq:New-charge}
\Delta\rts{1}{}{\xt}&\equiv& \rts{1}{q}{\xt}-\rts{1}{p}{\xt}\nonumber\\
\Delta\rts{2}{}{\xt}&\equiv& \rts{2}{p}{\xt}-\rts{2}{q}{\xt}\nonumber\\
\Delta\mu_{A_{1}}^2(\rt)&\equiv&
\mu_{A_{1}}^2(y_q,\rt)-\mu_{A_{1}}^2(y_p,\rt)\nonumber\\ 
\Delta\mu_{A_{2}}^2(\rt)&\equiv&
\mu_{A_{2}}^2(y_p,\rt)-\mu_{A_{2}}^2(y_q,\rt)\; .
\end{eqnarray}
Note that because of our choice $y_q > y_p$, $\Delta\mu^2$ as defined is
always positive\footnote{$\mu_{A}^2$ is proportional to the saturation
  scale, and therefore increases as one evolves away from the
  fragmentation region of a projectile.}.

Because the Green's functions in \eq{\ref{eq:Green-Gauss}} are
expressed naturally as Gaussians in the new variables introduced in \eq{\ref{eq:New-charge}}, we
can rewrite our general expression for the double inclusive
distribution as
\begin{eqnarray}
\left< \frac{\ud N_2}{\ud^2\pp \ud y_p \ud^2\qp \ud y_q}\right>_{_{\rm LLog}}
&=&
\int \left[D\rts{1}{p}{\xt} D\rts{2}{q}{\xt} 
D\Delta\rts{1}{}{\xt} D\Delta\rts{2}{}{\xt} \right]
\nonumber\\
&&\times Z_{y_p}[\rtv{1}{p}]\,G_{y_p,y_q}[\rtv{1}{p},\rtv{1}{q}]\,
Z_{y_q}[\rtv{2}{q}]\,G_{y_q,y_p}[\rtv{2}{q},\rtv{2}{p}]\nonumber\\
&&\times
\left.\frac{\ud N_1\left[\rtv{1}{p},\rtv{2}{p}\right]}{\ud^2\pp \ud y_p}\right|_{_{\rm LO}}
\left.\frac{\ud N_1\left[\rtv{1}{q},\rtv{2}{q}\right]}{\ud^2\qp \ud y_q}\right|_{_{\rm LO}}\; .
\label{eq:double-inclusive-2}
\end{eqnarray}
With the $Z_y$'s from \eqpar(\ref{eq:Z1}) and the $G_{y_q,y_p}$'s from
\eqpar(\ref{eq:gaussian-G}), the only ingredient missing in obtaining a
final expression for \eqpar(\ref{eq:single-inclusive-1}) and
\eqpar(\ref{eq:double-inclusive-2}) is the expression of the leading
order single inclusive spectrum in terms of the color charge densities
of the two nuclei. This expression and the subsequent simplification
of our equations for the inclusive distributions will be discussed in
the next section.

\begin{figure}[htbp]
\begin{center}
\includegraphics[scale=1]{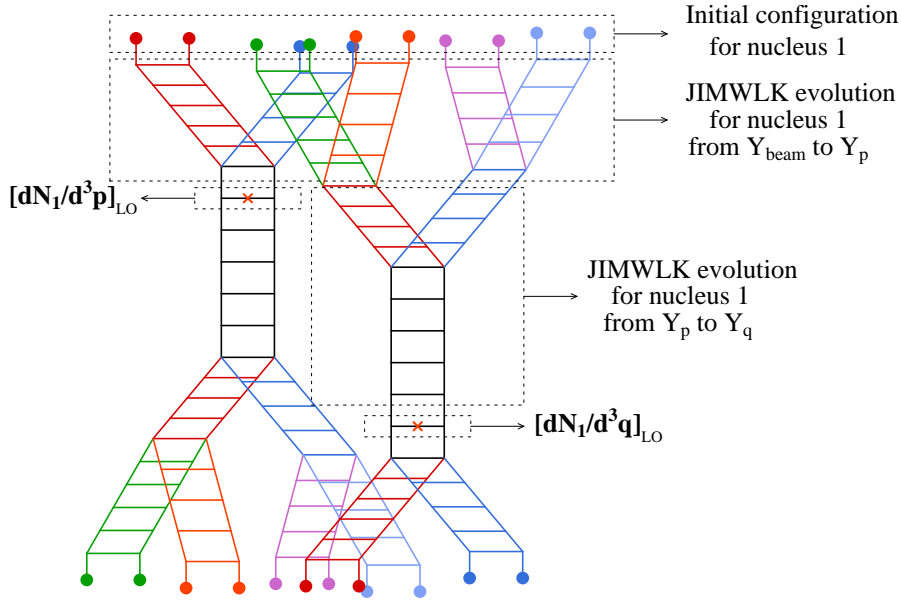}
\end{center}
\caption{\label{fig:2-gluon} Diagrammatic representation of the various building blocks in the
factorized formula for the inclusive 2-gluon spectrum. As in the previous figure,
the corresponding evolution from nucleus 2 at the bottom of the figure is not
shown explicitly.}
\end{figure}

\section{Gluon correlations from unintegrated gluon distributions}
\label{sec:ktfact}

The leading order single particle inclusive distribution, for a fixed distribution of sources, is given by 
\begin{eqnarray}
\left.\frac{\ud N_1\left[\rtv{1}{},\rtv{2}{}\right]}{\ud^2\pp \ud y_p}\right|_{_{\rm LO}}
 &=&\frac{1}{16\pi^3}
\lim_{x_0,y_0\to+\infty}\int \ud^3{\bf x}\, \ud^3{\bf y}
\;e^{ip\cdot(x-y)}
\;(\partial_x^0-iE_p)(\partial_y^0+iE_p)
\nonumber\\
&&\qquad\qquad\times\sum_{\lambda,a}
\epsilon_\lambda^\mu({\bf p})\epsilon_\lambda^\nu({\bf p})\;
A_\mu^a(x) [\rho_1,\rho_2]\; A_\nu^a(y)[\rho_1,\rho_2]\; .
\label{eq:single-inclusive-2}
\end{eqnarray}
The gauge fields $A_\mu^a(x) [\rho_1,\rho_2]$ are solutions of the
classical Yang-Mills equations in the forward light cone after the
nuclear collision for a fixed configuration of sources $\rho_{1,2}^a$ in
each of the nuclei. For Fourier modes $\kt$ of the color charge
densities which obey ${\wt\rho}_{1,2}(\kt) /{\bf k}_\perp^2 \gtrsim 1$ (which is
the case for $\qs \gtrsim k_\perp$), only numerical solutions for
$A_\mu(x)$ are
known~\cite{KrasnV1,KrasnV2,KrasnV3,KrasnNV1,KrasnNV2,KrasnNV3,Lappi1,Lappi2,Lappi3}.
However, for ${\wt\rho}_{1,2}/{\bf k}_\perp^2 \ll 1$, valid for $\qs \ll
k_\perp$, one can perturbatively expand the gauge field in powers of
${\wt\rho}_{1,2}/{\bf k}_\perp^2$ and one obtains~\cite{KovneMW1,KovchR1}
\begin{equation}
  p^2A^{\mu}_a({\bf p})=-if_{abc}\;g^3\int\frac{\ud^2\kp}{(2\pi)^2}\,L^\mu({\bf p},\kp)\frac{\tilde{\rho}^b_1(\kp)\tilde{\rho}_2^c(\pp-\kp)}{\kp^2\left(\pp-\kp\right)^2}
\label{eq:gauge-field}
\end{equation}
where $f_{abc}$ are the $SU(3)$ structure constants and
$\tilde{\rho}_{1,2}$ are the Fourier transforms of the color charge
densities of the two nuclei. Here $L^\mu$ is the
well-known~\cite{KuraeLF1,BalitL1} Lipatov vertex\footnote{The
    components of the Lipatov four vector are $L^+(\p,\k_\perp) =
    -\frac{\k_\perp^2}{p^-}$, $L^-(\p,\k_\perp) =
    \frac{(\p_\perp-\k_\perp)^2-\p_\perp^2}{p^+}$, $L^i(\p,\k_\perp) =
    -2\, \k_\perp^i$.}.

For the single inclusive distribution in
\eqpar(\ref{eq:single-inclusive-1}), using
\eqpar(\ref{eq:single-inclusive-2}) and \eqpar(\ref{eq:gauge-field}) and
the correlator
\begin{equation}
\left< {\wt\rho}^a(\kt){\wt\rho}^b ({\bf k}_\perp^\prime)\right> =
(2\pi)^2 \mu_{A}^2(y) \,\delta^{ab} \delta(\kt-{\bf k}_\perp^\prime)\; ,
\end{equation}
one obtains
\begin{eqnarray}
\left<\frac{\ud N_1}{\ud^2\pp \ud y_p}\right>_{_{\rm LLog}}
=
S_\perp\frac{2g^6 N_c(N_c^2-1)}{(2\pi)^5}
\frac{1}{\pp^2}
\int \ud^2\kt 
\frac{\mu^2_{A_1}(y_p,\kt)\mu^2_{A_2}(y_p,\pt-\kt)}
{{\bf k}_\perp^2(\pt-\kt)^2}\; ,
\label{eq:single-inclusive-3}
\end{eqnarray}
where $S_\perp$ is the transverse area of the overlap between the two
nuclei.  The unintegrated gluon
distribution can be expressed as~\cite{BlaizGV1,FujiiGV1,FujiiGV2}
\begin{equation}
  \phi_{_{A_{1,2}}}(x,k_\perp)\equiv
  \frac{\pi R_{_{A_{1,2}}}^2 {\bf k}_\perp^2}{4\alpha_s N_c} 
\int \ud^2\xt\; e^{i \kt \cdot \xt } \; 
\left<{\rm Tr}\left(U^{\dagger}(0) U(\xt)\right)\right> \; ,
\label{eq:unintBGV}
\end{equation}
where the matrices $U$ are adjoint Wilson lines evaluated in the
classical color field created by a given partonic configuration of the
nuclei $A_1$ or $A_2$. For a nucleus moving in the $-z$ direction,
\begin{eqnarray}
U(\x_\perp)\equiv \mathrm{P}_+ \exp\left[ig\int\limits_{-\infty}^{+\infty}
\ud z^+ \frac{1}{{\bf\nabla}_{\perp}^2}\,\rho_a(z^+,\x_\perp) T^a\right] \; .
\label{eq:pathGlue}
\end{eqnarray}
Here the $T^a$ are the generators of the adjoint representation of
$SU(N_c)$ and $\mathrm{P}_+$ denotes path ordering along the $z^+$
axis.  At large $\kt$, the Wilson lines can be expanded in powers of
the sources to give, for Gaussian correlations,
\begin{equation}
\phi_{A}(y,\kp)=g^2\pi (\pi R_{A}^2) (N_c^2-1) \frac{\mu_{A}^2(y,\kp)}{\kp^2}\; .
\label{eq:unint-gluon-dist}
\end{equation}
Substituting this relation in \eqpar(\ref{eq:single-inclusive-3}), we
obtain the well known $\kp$-factorization
expression~\cite{KovchM3,Braun1} for the single inclusive gluon
distribution valid for $p_\perp \gg \qs$:
\begin{equation}
\left<\frac{\ud N_1}{\ud^2\pp \ud y_p}\right>_{_{\rm LLog}}
=\frac{2\alpha_s N_c S_\perp}{2\pi^4 (N_c^2-1)}\frac{1}{\pp^2}
\int \frac{\ud^2\kt}{(2\pi)^2} 
\Phi_{A_1}(y_p,\kt)\Phi_{A_2}(y_p,\pt-\kt)\; ,
\label{eq:single-inclusive-4}
\end{equation}
where we denote $\Phi_A\equiv \phi_A/(\pi R_A^2)$ to be the unintegrated gluon
distribution {\sl per unit of transverse area}.

The corresponding expression for the double inclusive distribution is
more involved.  The r.h.s of \eqpar(\ref{eq:double-inclusive-2}) has the
product of two single inclusive distributions, one for a gluon with
three momentum ${\bf p}$ and likewise another for a gluon with three
momentum ${\bf q}$. From \eqpar(\ref{eq:single-inclusive-2}), this
corresponds to the product of four gauge fields. As for the single
inclusive case, the double inclusive gluon spectrum can be computed
numerically using lattice techniques where Yang-Mills equations are
solved to obtain the gauge fields as a function of proper time after
the collision. This computation has been carried out recently for the
Gaussian MV model~\cite{LappiSV1}. Because this model does not include
the effects of small $x$ evolution, it is not ideal for the purpose of
investigating the dynamics of long range rapidity correlations.
Incorporating small $x$ evolution effects in the non-perturbative
computation is outside the scope of the present work. We will instead
consider here, as in the previous discussion of the single inclusive
distribution--see \eq{\ref{eq:gauge-field}}, the perturbative limit of
$p_\perp$, $q_\perp\gg \qs$, where the gauge fields can be expanded as
bilinear products of the color sources of the two nuclei. The
dependence of the leading order double inclusive gluon spectrum on
four gauge fields, then translates, in this perturbative limit to the
product of eight color charge densities. The averages over color
sources in \eqpar(\ref{eq:double-inclusive-2}) are therefore averages
over the general matrix element
\begin{eqnarray}
\mathcal{F}^{bcdefghi}({\bf p},{\bf q};\{\kpn{i}\})
&\equiv&
\Big<
{\wt\rho}^{*f,p}_1(\kpn{2})\;{\wt\rho}^{*h,q}_1(\kpn{4}) 
{\wt\rho}^{b,p}_1(\kpn{1})\;{\wt\rho}^{d,q}_1(\kpn{3}) \nonumber\\
&&
\!\!\!\!\times
{\wt\rho}^{*g,p}_2(\pp-\kpn{2})\;{\wt\rho}^{*i,q}_2(\qp-\kpn{4}) 
{\wt\rho}^{c,p}_2(\pp-\kpn{1})\; {\wt\rho}^{e,q}_2(\qp-\kpn{3}) 
\Big>\;,
\nonumber\\
&&
\label{eq:F}
\end{eqnarray}
where we denote by a superscript $p$ or $q$ the rapidity which 
the color sources correspond to.
Further, these products of gauge fields contain bi-linear scalar
products of the Lipatov vertices. These can be
simplified~\cite{DumitGMV1,DusliFV1} and expressed as
\setlength\arraycolsep{.1pt}
\begin{eqnarray}
\mathcal{G}({\bf p},{\bf q};\{\kpn{i}\})
=
\frac{16}{(2\pi)^8}
\;
&&\frac{\big[
\big(\kpn{1}\cdot \pp-\kpn{1}^2\big)
\big(\kpn{2}\cdot \pp-\kpn{2}^2\big)
+
\big(\kpn{1}\times\pp\big)
\cdot
\big(\kpn{2}\times\pp\big)
\big]}
{\kpn{1}^2\kpn{2}^2\pp^2
\big(\pp-\kpn{1}\big)^2
\big(\pp-\kpn{2}\big)^2
}
\nonumber\\
\quad\times
&&\frac{\big[
\big(\kpn{3}\cdot \qp-\kpn{3}^2\big)
\big(\kpn{4}\cdot \qp-\kpn{4}^2\big)
+
\big(\kpn{3}\times\qp\big)
\cdot
\big(\kpn{4}\times\qp\big)
\big]
}
{
\kpn{3}^2\kpn{4}^2\qp^2
\big(\qp-\kpn{3}\big)^2
\big(\qp-\kpn{4}\big)^2}\; .
\nonumber\\
&&
\label{eq:Lipatov-product}
\end{eqnarray}
\setlength\arraycolsep{1.4pt}

The double inclusive distribution in
\eqpar(\ref{eq:double-inclusive-2}), for  transverse momenta $p_\perp, q_\perp
\gg \qs$, can therefore be expressed as
\begin{eqnarray}
\left< 
\frac{\ud N_2}{\ud^2\pp \ud y_p \ud^2\qp \ud y_q}\right>_{_{\rm LLog}}
&=&
\frac{g^{12}}{16(2\pi)^6}
\,
f^{abc}f^{a^\prime de}f^{afg}f^{a^\prime hi}
\nonumber\\
&&
\!\!\!\!\times
\int\prod_{i=1}^4 \ud^2\kpn{i}\;
\mathcal{G}({\bf p},{\bf q};\{\kpn{i}\})\,
\mathcal{F}^{bcdefghi}({\bf p},{\bf q};\{\kpn{i}\})\; ,
\label{eq:double-inclusive-3}
\end{eqnarray}
in terms of $\mathcal{F}^{bcdefghi}({\bf p},{\bf q};\{\kpn{i}\})$ and
$\mathcal{G}({\bf p},{\bf q};\{\kpn{i}\})$ defined above. 

We shall now sketch how one evaluates these quantities with further details of
the computation given in appendix A. We begin with
the evaluation of the color averages in $\mathcal{F}^{bcdefghi}({\bf
  p},{\bf q};\{\kpn{i}\})$.  Because the $Z$'s in \eqpar(\ref{eq:Z1})
and the $G$ in \eqpar(\ref{eq:Green-Gauss}) are Gaussian weight
functionals, the relevant color source correlators are the equal rapidity correlators 
\begin{eqnarray}
\left<\wt\rho^{*a,p}_{1}(\kp)\wt\rho^{b,p}_{1}(\kp^\prime)\right>
&=&
(2\pi)^2 \delta^{ab}\,\delta^2(\kp-\kp^\prime)\,
\mu^2_{A_{1}}(y_p,\kp)\; ,
\nonumber\\
\left<\wt\rho^{*a,q}_{2}(\kp)\wt\rho^{b,q}_{2}(\kp^\prime)\right>
&=&
(2\pi)^2 \delta^{ab}\,\delta^2(\kp-\kp^\prime)\,
\mu^2_{A_{2}}(y_q,\kp)\; ,
\nonumber\\
\left<\Delta\wt\rho^{*a}_{1,2}(\kp)\Delta\wt\rho^{b}_{1,2}(\kp^\prime)\right>
&=&
(2\pi)^2 \delta^{ab}\,\delta^2(\kp-\kp^\prime)\,
\Delta\mu^2_{A_{1,2}}(\kp)\; ,
\end{eqnarray}
and the non-equal rapidity correlators\footnote{This follows from the
  vanishing of terms odd in $\rho$ or $\Delta\rho$.}
\begin{eqnarray}
\left<\wt\rho^{*a,q}_{1}(\kp)\wt\rho^{b,p}_{1}(\kp^\prime)\right>
&=&
\left<
\left(\Delta\wt\rho_1^*(\kp)+\wt\rho^{*a,p}_{1}(\kp)\right)
\wt\rho^{b,p}_{1}(\kp^\prime) 
\right>
\nonumber\\
&=&
(2\pi)^2 \delta^{ab}\,\delta^2(\kp-\kp^\prime)\,
\mu^2_{A_{1}}(y_p,\kp)\; ,\nonumber\\
\left<\wt\rho^{*a,q}_{2}(\kp)\wt\rho^{b,p}_{2}(\kp^\prime)\right>
&=&
\left<
\wt\rho^{*a,q}_{2}(\kp) 
\left(\Delta\wt\rho_2(\kp^\prime)+\wt\rho^{b,q}_{2}(\kp^\prime)\right) 
\right>
\nonumber\\
&=&
(2\pi)^2 \delta^{ab}\,\delta^2(\kp-\kp^\prime)\,
\mu^2_{A_{2}}(y_q,\kp)\; . 
\end{eqnarray}
The correlators for the dependent variables $\rho^q_1$ and $\rho^p_2$
(see \eqpar(\ref{eq:New-charge})) are
\begin{eqnarray}
\left<\wt\rho^{*a,q}_{1}(\kp)\wt\rho^{b,q}_{1}(\kp^\prime)\right>
&=&
(2\pi)^2 \delta^{ab}\,\delta^2(\kp-\kp^\prime)\,
\mu^2_{A_{1}}(y_q,\kp)\; ,\nonumber\\
\left<\wt\rho^{*a,p}_{2}(\kp)\wt\rho^{b,p}_{2}(\kp^\prime)\right>
&=&
(2\pi)^2 \delta^{ab}\,\delta^2(\kp-\kp^\prime)\,
\mu^2_{A_{2}}(y_p,\kp)\; .
\end{eqnarray}

With the relations listed, we can now evaluate \eqpar(\ref{eq:F}).
Simple combinatorics gives us a total of 9 possible pairwise
contractions in \eqpar(\ref{eq:F}). The details of these are listed in
appendix A. One of the contributions
(\eqpar(\ref{eq:F-9})) gives the non-correlated contribution to the two
gluon inclusive distribution. Subtracting this term therefore results
in the correlated two gluon inclusive distribution
\begin{eqnarray}
C(\p,\q) 
\equiv
\left<\frac{\ud N_2}{\ud y_p \ud^2\pp \ud y_q \ud^2\qp}\right> 
 - 
\left<\frac{\ud N}{\ud y_p \ud^2\pp}\right> 
\left<\frac{\ud N}{\ud y_q \ud^2\qp}\right>\; .
\label{eq:C-def}
\end{eqnarray}
When one evaluates the other 8 terms that contribute to $C(\p,\q)$,
one observes that only 4 of these give leading contributions. The
$\delta$-function contributions from these terms (eqs.~(\ref{eq:F-1}),
(\ref{eq:F-2}), (\ref{eq:F-3}), (\ref{eq:F-6})) give $\kpn{1}=\kpn{2}$
and $\kpn{3}=\kpn{4}$. Substituting this into the expression for
$\mathcal{G}({\bf p},{\bf q};\{\kpn{i}\})$ in
\eqpar(\ref{eq:Lipatov-product}), one finds that it simplifies
considerably to read
\begin{equation}
\mathcal{G}({\bf p},{\bf q};\{\kpn{i}\})
=
\frac{16}{(2\pi)^8\, \kpn{1}^2\kpn{3}^2\pp^2\qp^2\left(\pp-\kpn{1}\right)^2\left(\qp-\kpn{3}\right)^2}
\end{equation}
Combining the four leading contributions from
$\mathcal{F}^{bcdefghi}({\bf p},{\bf q};\{\kpn{i}\})$ and the
corresponding expressions from $\mathcal{G}({\bf p},{\bf
  q};\{\kpn{i}\})$ to $C({\bf p},{\bf q})$ (see eqs.~(\ref{eq:C-1}),
(\ref{eq:C-3}), (\ref{eq:C-2}) and (\ref{eq:C-6}) in
appendix A),  we obtain
\setlength\arraycolsep{.4pt}
\begin{eqnarray}
C({\bf p},{\bf q})
=&&
\frac{\alpha_s^{2}}{16 \pi^{10}}
\frac{N_c^2(N_c^2-1) S_\perp}{d_A^4\; \pp^2\qp^2} \int \ud^2\kpn{1}\times
\nonumber\\
&\bigg\{
&\Phi_{A_1}^2(y_p,\kpn{1})\Phi_{A_2}(y_p,\pp-\kpn{1})
\left[
\Phi_{A_2}(y_q,\qp+\kpn{1})
+
\Phi_{A_2}(y_q,\qp-\kpn{1})
\right]
\nonumber\\
&+&\Phi_{A_2}^2(y_q, \kpn{1})\Phi_{A_1}(y_p,\pp-\kpn{1})
\left[
\Phi_{A_1}(y_q,\qp+\kpn{1})
+
\Phi_{A_1}(y_q,\qp-\kpn{1})
\right]\!
\bigg\}\, ,
\nonumber\\
&&
\label{eq:double-inclusive-4}
\end{eqnarray}
\setlength\arraycolsep{1.4pt}
where the $\Phi$'s are unintegrated gluon distributions per unit of
transverse area and $d_A = N_c^2-1$. We have used here the relation
between $\mu^2$ and the unintegrated gluon distribution $\Phi$ given
in \eqpar(\ref{eq:unint-gluon-dist}).  This expression is the central
result of this paper\footnote{We note that expressions for the double
  inclusive cross-section have been previously derived~\cite{Braun2}
  within the framework of Local Reggeon Field Theory~\cite{CiafaM1}.
  At present the connection between the two frameworks is completely
  unclear.}. We have obtained an expression for the double inclusive
gluon distribution, valid to all orders in perturbation theory to
leading logarithmic accuracy in $x$ and for momenta $p_\perp,q_\perp \gg
\qs$, entirely in terms of the unintegrated gluon distributions of the
two nuclei evaluated at the rapidities $y_p$ and $y_q$ where $y_p <
y_q$. The corresponding expression for $y_p > y_q$ is obtained by
replacing $A_1 \leftrightarrow A_2$ and $y_{p,q}\rightarrow -y_{p,q}$.  We should emphasize that the notation used in \eqpar(\ref{eq:double-inclusive-4}) stipulates that the un-integrated gluon distributions are evaluated at rapidities $y_{p,q}\pm Y_{\rm {beam}}$.

\section{Running coupling BK evolution}
\label{sec:BK}

In the previous section, we established that the correlated two gluon
spectrum for arbitrary rapidities can be computed in terms of the
unintegrated gluon distributions of the two nuclei evaluated at these
rapidities. In this section, we shall discuss how one computes this
unintegrated gluon distribution and its evolution with $x$. In the
next section, we shall use the results for the unintegrated gluon
distribution to evaluate \eqpar(\ref{eq:double-inclusive-4}) for the
correlated inclusive two gluon distribution.

In \eqpar(\ref{eq:unintBGV}), we defined the unintegrated gluon
distribution in a nucleus in terms of the correlator of two adjoint
Wilson lines averaged over the color charge distribution in a nucleus.
Because these averages $\big<\cdots\big>$ and those of correlators of
fundamental Wilson lines are Gaussian correlators in the large $N_c$
limit, one can express these correlators respectively
as~\cite{BlaizGV1,BlaizGV2}
\begin{eqnarray}
&&
{\rm Tr}\left<U(0)U^\dagger(\r_\perp)\right>_{_Y}
= N_c^2 \; e^{-C_{_A}\Gamma(r_\perp,Y)} \nonumber \\
&& {\rm Tr}\left<{\wt U}(0){\wt U}^\dagger(\r_\perp)\right>_{_Y}
=N_c \; e^{-C_{_F}\Gamma(r_\perp,Y)}\; ,
\label{eq:largeN-corr}
\end{eqnarray}
where $C_{_A}= N_c$ is the Casimir in the adjoint representation and
$C_{_F}= (N_c^2-1)/2 N_c$ is the Casimir in the fundamental
representation. The function $\Gamma$ is closely
related~\cite{FujiiGV2} to the variance of the non-local Gaussian
weight functional in \eqpar(\ref{eq:Z1}) and is therefore the same in
both the fundamental and adjoint cases. One can therefore, in the
large $N_c$ limit, simply express the correlator of two adjoint Wilson
lines as the square of the correlator of two fundamental Wilson
correlators.

The correlator of two Wilson lines in the fundamental representation
is simply related to the dipole amplitude for the scattering of a
quark-antiquark dipole (of transverse separation ${\bf r}_\perp$) off
a nucleus as\footnote{We assume translation invariance in the
  transverse plane to set the quark transverse coordinate to zero.}
 \begin{equation}
 T(\r_\perp,Y)
=
1- \frac{1}{N_c}{\rm Tr} \left<
 {\tilde U}^\dagger(0) {\tilde U}(\r_\perp)\right>_{_Y} \; ,
 \label{eq:dipole}
 \end{equation}
 where ${\tilde U}$ is a Wilson line in the fundamental
 representation. (See our previous discussion of these in the context
 of \eqpar(\ref{eq:BK-Factorization}).)  Using
 \eqpar(\ref{eq:largeN-corr}), one can write the unintegrated gluon
 distribution in the adjoint representation (per unit of transverse area) in \eqpar(\ref{eq:unintBGV}) as
\begin{equation}
\Phi_{{A_{1,2}}}(x,k_\perp) 
= 
\frac{\pi N_c k_\perp^2}{2\,\alpha_s}
\int\limits_0^{+\infty} r_\perp \ud r_\perp 
\;
J_0(k_\perp r_\perp)
\,
\left[1-T_{{A_{1,2}}}(r_\perp,\ln(1/x))\right]^2
 \; .
\label{eq:phiadjoint}
\end{equation}

We therefore need to determine the dipole amplitude $T$ and its
evolution with rapidity $Y$ (=$\ln(x_0/x)$) as an input in
\eq{\ref{eq:phiadjoint}} to extract the unintegrated gluon
distribution. The dipole amplitude is obtained from the
Balitsky-Kovchegov (BK) equation~\cite{Balit2,Kovch2}, which is a
non-linear evolution equation describing both gluon emission and
multiple scattering effects in the interaction of the quark-antiquark
dipole with a nucleus in the large $N_c$ limit. It can be
expressed as
\setlength\arraycolsep{.1pt}
\begin{eqnarray}
\frac{\partial T(\r,Y)}{\partial Y} 
= 
\int \ud &&\rb{1}\; {\mathcal K}_{_{\rm LO}}(\r,\rb{1},\rb{2})\times\;\nonumber\\
&&\big[ T(\rb{1},Y) + T(\rb{2},Y) - T(\r,Y) - T(\rb{1},Y)\,T(\rb{2},Y)\big] \; ,
\label{eq:BK-LO}
\end{eqnarray}
\setlength\arraycolsep{1.4pt}
with the leading order BFKL kernel~\cite{Lipat2} given by
\begin{equation}
{\mathcal K}_{_{\rm LO}}(\r,\rb1 ,\rb2) 
= 
\frac{\alpha_s N_c}{2\pi^2}\, \frac{\r^2}{ \rb1^2 \rb2^2} \; ,
\label{eq:BFKL-kernel}
\end{equation}
where $\rb2 \equiv \r - \rb1$.  As we discussed previously, the BK
equation for the amplitude is equivalent to the corresponding JIMWLK
equation~\cite{JalilKMW1,JalilKLW1,JalilKLW2,JalilKLW3,JalilKLW4,IancuLM1,IancuLM2,FerreILM1}
of the Color Glass Condensate, in a mean field (large $N_c$) 
approximation where higher order dipole correlators are
neglected.

In the context of the BK equation, the leading order kernel
corresponds to resumming the leading $(\alpha_s \ln(x_0/x))^n$ terms
arising at small $x$ from all orders in perturbation theory. It is
well known however that running coupling contributions qualitatively
modify the small $x$ evolution beyond leading logarithms in $x$ and
there has been considerable recent work to include these
corrections to the BK
equation~\cite{BalitC1,KovchW1}.  The running coupling equation
describing the evolution of the dipole amplitude however takes exactly
the same form as \eq{\ref{eq:BK-LO}} with a modified evolution kernel
given by
\begin{equation}
{\mathcal K}_{\rm Bal.}(\rb,\rb{1},\rb{2}) = \frac{\alpha_s(\r) N_c}{\pi}\left[ \frac{\r^2}{\rb{1}^2 \rb{2}^2} +\frac{1}{\rb{1}^2}\left(\frac{\alpha_s(\rb{1}^2)}{\alpha_s(\rb{2}^2)}-1\right)+\frac{1}{\rb{2}^2}\left(\frac{\alpha_s(\rb{2}^2)}{\alpha_s(\rb{1}^2)}-1\right)\right] \, .
\label{eq:NLO-BFKL-kernel}
\end{equation}
The subscript in ${\mathcal K}_{\rm Bal.}$ refers to the ``Balitsky
prescription'' for the evolution kernel, which corresponds to a 
scheme where some particular ultra-violet finite terms are also included along with the
running coupling contributions to make the remainder numerically less
important.  For a more detailed discussion, we refer the reader
to~\cite{AlbacK1}.  In this work, the NLO contributions not
encompassed by the kernel in \eq{\ref{eq:NLO-BFKL-kernel}} will be
ignored. As argued previously~\cite{AlbacAMS1}, these contributions
are systematically smaller than the running coupling contribution
included here, especially at large rapidities.

\begin{figure}[htbp]
\begin{center}
\includegraphics[scale=0.6]{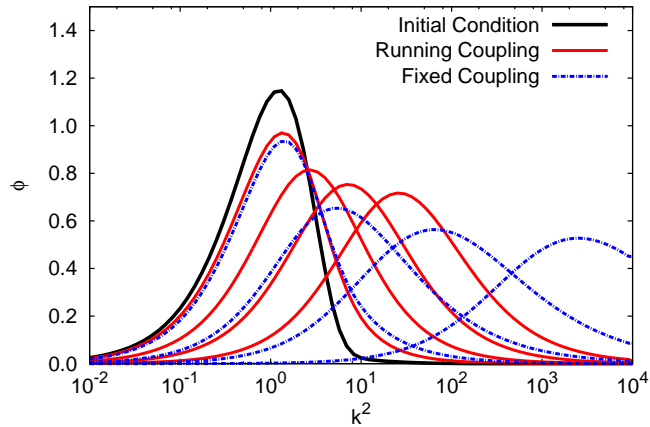}
\end{center}
\caption{Unintegrated gluon distribution in the adjoint representation
 at $Y=0,2,6,10,15$ (from left curve rightwards) with the Balitsky
 prescription for the kernel in \eq{\protect\ref{eq:NLO-BFKL-kernel}} 
 as well as for the fixed coupling case. The distribution is in units 
 of $N_c \pi R_A^2/\alpha_s$.} 
\label{fig:wave}
\end{figure}

In \fig\ref{fig:wave}, we show results for the unintegrated gluon
distribution versus transverse momentum squared determined from the
evolution with rapidity of the dipole amplitude in the adjoint
representation (see \eq{\ref{eq:phiadjoint}}) with i) the fixed
coupling BK kernel, and ii) with the Balitsky prescription for the
kernel in \eq{\ref{eq:NLO-BFKL-kernel}}).  As we will describe below,
the initial conditions for the latter figure are constrained by
fixed target e+A data. We note that the evolution of the unintegrated
gluon distribution with Balitsky's prescription for the running
coupling effects is significantly slower than the evolution with a
fixed coupling constant.

\begin{figure}
\begin{center}
\includegraphics[scale=.55]{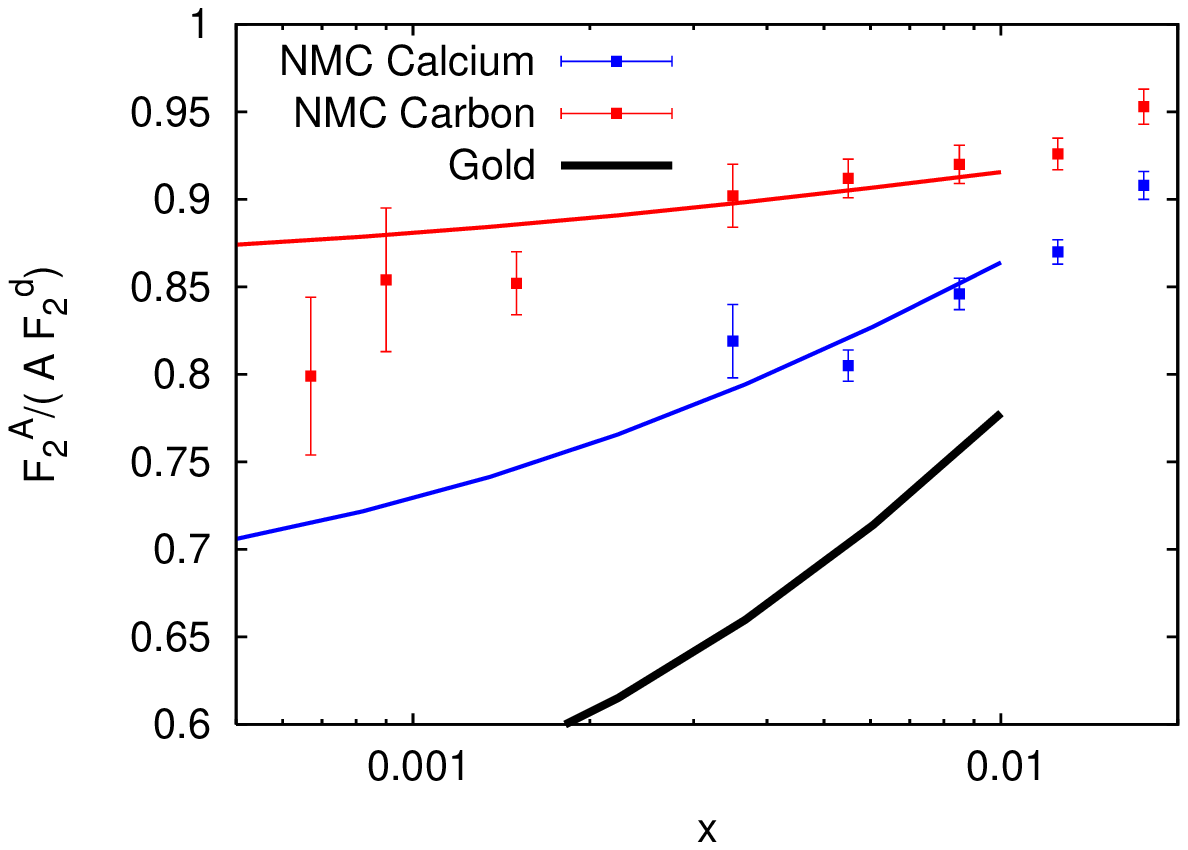}
\includegraphics[scale=.55]{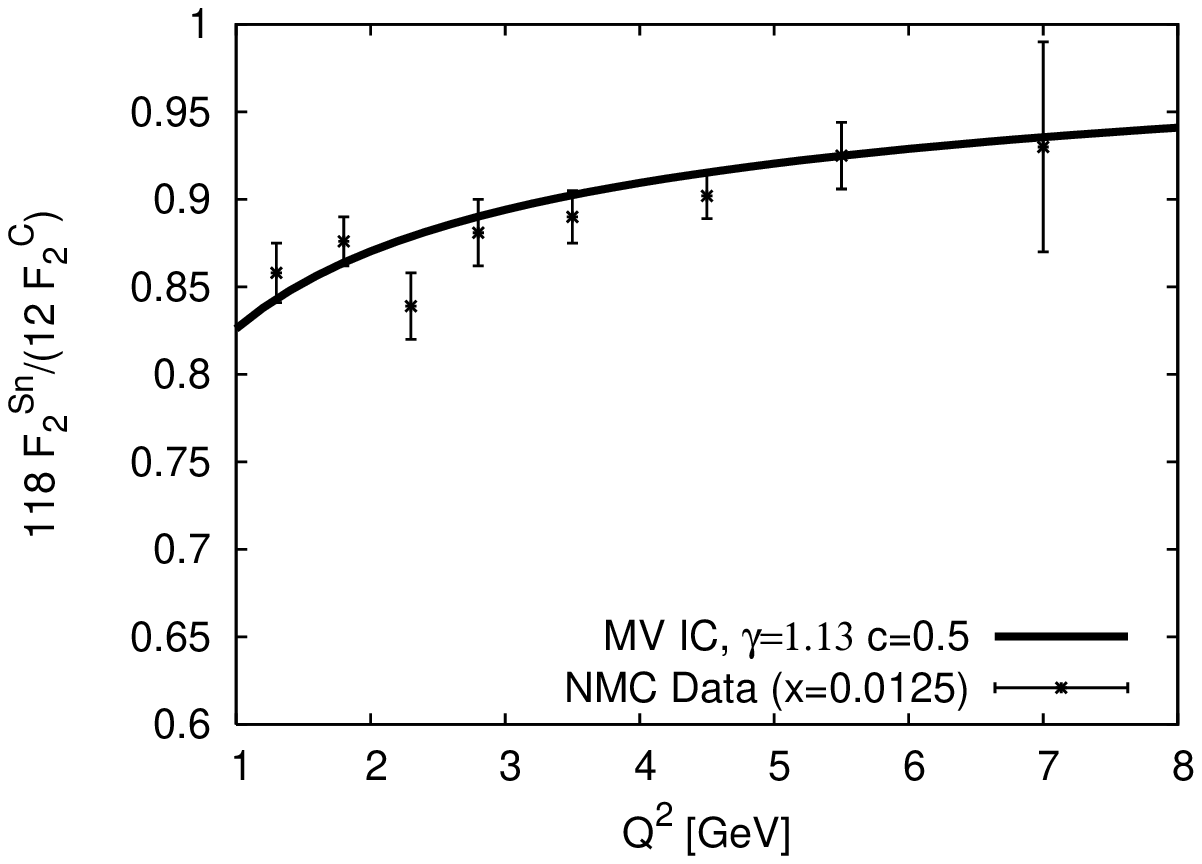}
\end{center}
\caption{The $x$ and $Q^2$ dependence of the normalized ratio of
  structure functions $F_2$ in nuclei. The curves in the left figure
  includes effects due to the small $x$ evolution of the dipole
  cross-section described by the BK evolution with the modified kernel
  in \eq{\protect\ref{eq:NLO-BFKL-kernel}}.  The curve in the right
  figure is sensitive to the $Q^2$ dependence of the initial condition
  alone because it is evaluated at relatively large $x$. Details
  regarding the parameters of the initial condition are discussed in
  appendix B. The data are from the NMC
  collaboration~\cite{AmaudA1}. }
\label{fig:NMC-xQ2}
\end{figure}

\begin{figure}
\begin{center}
\includegraphics[scale=.55]{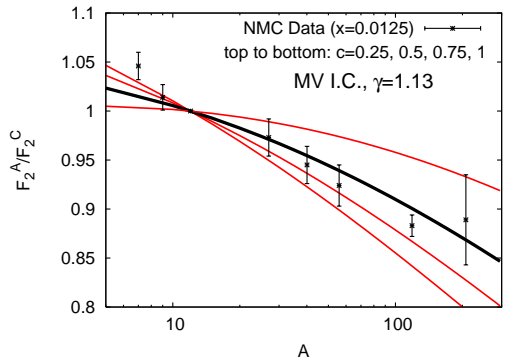}
\end{center}
\caption{The $A$ dependence of the ratio of structure functions given
  by data from the NMC collaboration~\cite{ArneoA1}. The corresponding
  curves for other initial conditions are in appendix B.}
\label{fig:NMCAg}
\end{figure}

The BK equation with the modified kernel in
\eqpar(\ref{eq:NLO-BFKL-kernel}) was first applied in
Refs.~\cite{AlbacAMS1,AlbacAMS2} to a phenomenological study of the
HERA data on the proton structure function $F_2$. Two sets of initial
conditions for the dipole amplitude at the initial rapidity $Y=Y_0$
were used--the GBW~\cite{GolecW1} and MV initial
conditions~\cite{McLerV1,McLerV2,McLerV3}--and their parameters
determined from fits to the HERA data. To constrain the initial
conditions for nuclei and therefore extract the nuclear unintegrated
gluon distribution, we performed a fit to the available NMC data on
the nuclear structure function $F_{2,A}(x,Q^2)$. The details of the
fit and the results are described in detail in appendix B. We show
here in \figs\ref{fig:NMC-xQ2} and \ref{fig:NMCAg} representative
plots of fits to $x$, $Q^2$ and $A$ dependence of the fixed target e+A
data. Good fits to the available data are obtained for both sets of
initial conditions for particular parameters. With the initial
conditions for the BK equation
fixed by the NMC data, we shall now use the corresponding unintegrated
gluon distribution to study long range rapidity correlations in the
Glasma.

\section{Results for long range rapidity correlations in the Glasma}
\label{sec:results}

In this section, we will make use our result in
\eq{\ref{eq:double-inclusive-4}} for the double inclusive gluon
distribution to compute long range rapidity correlations in A+A
collisions at RHIC and the LHC. The essential ingredient in
\eq{\ref{eq:double-inclusive-4}} is the unintegrated gluon
distribution which, as shown in \eq{\ref{eq:phiadjoint}}, is simply
related to the dipole scattering amplitude. The evolution of the
dipole scattering amplitude with rapidity (or $x$) is described by the
BK evolution equation given in \eq{\ref{eq:BK-LO}}, with the modified kernel
given in \eq{\ref{eq:NLO-BFKL-kernel}}. The rapidity dependence of the
double inclusive gluon spectrum therefore provides a sensitive test of
high energy QCD evolution.

Equation~\nr{eq:double-inclusive-4} is derived
in the leading $\ln x$ approximation, where all transverse momenta
are assumed to be parametrically of the same order
as $\qs$.  In this approximation
the $x$-values at which the unintegrated gluon distributions are evaluated
are not exactly determined, as long as $x \sim e^{\pm y} \qs/\sqrt{s} $,
where $y$ is the appropriate rapidity of the produced gluon ($y_p$ or $y_q$)
and the sign depends on the nucleus (1 or 2) considered.
We define the longitudinal momentum fractions of the
produced gluons with respect to nucleus 1 or 2 (denoted by subscripts)
\begin{eqnarray}
x_{1p}&=& \frac{p_\perp}{\sqrtv{s}}e^{-y_p}\,\,\,;\,\,\,
x_{1q}= \frac{q_\perp}{\sqrtv{s}}e^{-y_q}\nonumber \\
x_{2p}&=& \frac{p_\perp}{\sqrtv{s}}e^{+y_p}\,\,\,;\,\,\,
x_{2q}= \frac{q_\perp}{\sqrtv{s}}e^{+y_q}
\label{eq:x}
\end{eqnarray}
In the above expression, $p_\perp$ and $q_\perp$ are the transverse momenta
of the produced gluons. The unintegrated gluon distributions 
with momentum argument $\pp\pm\kpn{1}$ and $\qp\pm\kpn{1}$ in 
\eq{\ref{eq:double-inclusive-4}} are evaluated at these values of 
the momentum fraction. For the unintegrated distribution
with momentum argument $\kpn{1}$ we replace the transverse
momentum in \eq{\ref{eq:x}} by $(\pp + \qp )/2$ to make
our evaluation of \eq{\ref{eq:double-inclusive-4}} manifestly
symmetric in $\pp$  and $\qp$\footnote{Another option would be to replace the momentum in \eq{\ref{eq:x}} by $\qs(x)$.  We have tried this and found our results to be insensitive to the choice of scale.}. Our derivation in Sec.~\ref{sec:ktfact}
makes it clear that the term
with $\Phi^2$ in  \eq{\ref{eq:double-inclusive-4}} should be evaluated
at a rapidity scale that is the \emph{earlier} of the two rapidity scales
$y_p$ and $y_q$ in the evolution of the corresponding nucleus. This 
prescription guarantees that the same is true when the scale is parametrized 
in terms of $x$ instead of rapidity.

The solution of the BK equation is reliable when the gluon density is
large. The initial condition for the evolution is typically set at $x
\leq 0.01$.  For larger values of $x$, one expects the BK description
to break down; we use instead a phenomenological extrapolation (used
previously in~\cite{GelisSV1,FujiiGV2}) for the unintegrated gluon
distribution which has the form
\begin{eqnarray}
\phi(x,\kp)=\left( \frac{1-x}{1-x_0}\right)^\beta \phi(x_0,\kp)\,,
\label{extrap}
\end{eqnarray}
where $x_0=0.01$ and the parameter $\beta=4$.  This extrapolation to
large $x$ is unreliable and depends on physics which is not amenable
to the renormalization group approach advocated here.  However, in
experiments with finite kinematic reach, it is inevitable that one is
sensitive to the non-perturbative physics at large $x$ in some
kinematic range.  For example, from the kinematic expressions in
\eq{\ref{eq:x}}, the unintegrated gluon distribution of gluons having
$p_\perp=0.5$~GeV at RHIC energies of $\sqrtv{s}=200$~GeV/nucleon will
begin to be sensitive to the large $x$ extrapolation of the
distribution at $y_p\approx 1.4$ units in rapidity.  At the LHC
energy, the range in rapidity where we avoid this sensitivity is much
greater.  At $\sqrtv{s}=5.5$~TeV, the same gluon does not probe the
large $x$ extrapolation of the unintegrated gluon distribution until
$y_p=4.7$

With these caveats in mind, we shall now examine the two gluon
inclusive distributions in A+A collisions both at RHIC 
($\sqrt{s}=200$~GeV) 
and at the LHC ($\sqrt{s}=5.5$~TeV).  The beam rapidities for
these energies are $Y_{\rm {beam}}\approx \pm \ln\left(
  \frac{\sqrt{s}}{M_{\rm nucleon}}\right)\approx 5.36$ and $8.68$ for
RHIC and LHC respectively.  We will first consider RHIC collisions and
compare our results to recently measured long range rapidity
correlations in the near-side ridge by the PHOBOS collaboration.  The
experimental quantity of interest is $\frac{1}{N_{\rm
    {trig.}}}\frac{dN}{d\Delta\eta}$, where the trigger particle
consists of all particles having $p_\perp \geq 2.5$~GeV and an
acceptance in rapidity in the range $0 \leq \eta^{\rm{trig.}} \leq
1.5$.  The particles associated with this trigger have momenta larger
than 4 (35) MeV at a rapidity of 3 (0).  In performing the $\Delta
\eta$ projection in the experiment, the near side yield is integrated
over $\vert \Delta\phi_{pq}\vert \leq 1$. Hence in computing the
per-trigger yield, we should in principle also integrate our two
particle correlation $C(\p,\q)$ over the PHOBOS acceptance.  We will
instead perform a more qualitative comparison here by computing
instead our two particle correlation at representative values of the
trigger and associated particle momenta and multiplying the result by
the phase space volume corresponding to the PHOBOS acceptance.

\begin{figure}[h!]
\begin{center}
\includegraphics[scale=.55]{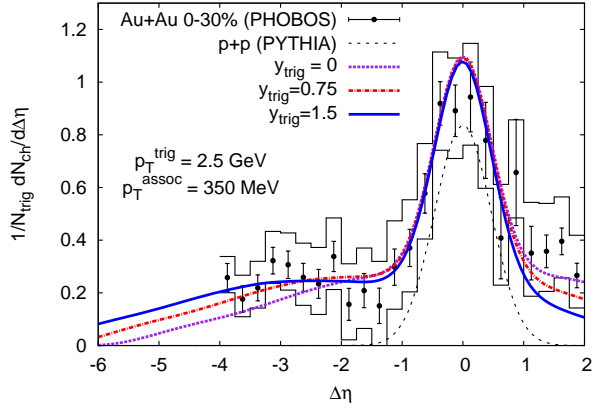}
\end{center}
\caption{Comparison of our results for long range rapidity
  correlations to data from the PHOBOS collaboration~\cite{AlverA1}.
  The curves shown are obtained by adding our result (expressed by
  \eq{\protect\ref{eq:C2-expt}} for long range rapidity correlations in the PHOBOS acceptance to the short range jet correlation in p+p collisions
  obtained using PYTHIA. 
}
\label{fig:PHOBOS-comp}
\end{figure}

For the trigger particle we take $p_\perp=2.5$~GeV at $y_{\rm
  {trig.}}=0, 0.75, $ and $1.5$ units in rapidity.  We assume the
associated particle has mean $p_\perp=350$~MeV.  For all cases, we
compute the yield at $\Delta\phi_{pq}=0$.  Then, in terms of our
expression for the two particle cumulant, the required quantity can be
written as
\begin{equation}
\frac{1}{N_{\rm {trig.}}}\frac{\ud N}{\ud \Delta\eta}
\approx \mathcal{V}_{ps}^{\rm{assoc.}} F(\Delta\phi_{pq}=0)
\frac{C(p_\perp^{\rm{trig.}}, p_\perp^{\rm{assoc.}}, y_{\rm{trig.}},
  y_{\rm{assoc.}}=y_{\rm{trig.}}+\Delta\eta,\Delta\phi_{pq}=0)}{\ud N_1(p_\perp^{\rm{trig.}},
  y_{\rm{trig.}})} \; ,
\label{eq:C2-expt}
\end{equation}
where $C(\p,\q)$ is the two particle cumulant given by
\eq{\ref{eq:double-inclusive-4}}.  In the above expression, the phase
space volume corresponding to the trigger particle cancels out; we are
left with an overall factor from the associated particle's phase space
volume, $\mathcal{V}_{ps}^{\rm{assoc.}}$, which we estimate to be
$\mathcal{V}_{ps}^{\rm{assoc.}}=\pi$~GeV$^2$. We arrive at this
estimate by performing the angular integration over
$\phi_{\rm{assoc.}}$ times the $p_\perp$ integration over the
acceptance.  Other than $\mathcal{V}_{ps}^{\rm{assoc.}}$, the only
additional parameter in our expression is $\alpha_s(\qs^2)$ which we
take to be $\alpha_s=0.35$.  With these stated values of $\alpha_s$
and $\mathcal{V}_{ps}^{\rm{assoc.}}$, our overall normalization is now
fixed.  The function $F(\Delta\phi_{pq})$ comes from the collimation
of the Glasma flux tubes due to radial flow as discussed in
Ref.~\cite{DumitGMV1}.  At $\Delta\phi_{pq}=0$, this can be expressed
as
\begin{equation} 
F(\Delta\phi_{pq}=0) = \cosh (\tanh^{-1}\beta)
\label{eq:flow}
\end{equation}
where $\beta=V/c$ is the radial flow velocity.

To take into account the short range correlation from fragmentation
not included in our formalism we add to  \eq{\ref{eq:C2-expt}}
the short range jet correlation resulting from PYTHIA.
The result is compared to the PHOBOS experimental
data~\cite{AlverA1} in \fig\ref{fig:PHOBOS-comp}. 
One can see that the agreement with data is quite good.  In principle the collimation from radial flow through \eq{\ref{eq:flow}} can be a function of rapidity.  We have estimated this effect by assuming that the space-time and momentum space rapidity are strongly correlated.  
From fits to BRAHMS data
\cite{Lee2,Murra2,Debbe2} on the inclusive hadron spectrum, we
estimate the $\eta$ dependence of the flow velocity to be $\beta(\eta)
= 0.72 - 0.04\vert \eta\vert$. When including this rapidity dependent flow through \eq{\ref{eq:flow}}, the effect is so small that it would not result in a 
visible change to the curves plotted in figure \ref{fig:PHOBOS-comp}.

\begin{figure}[h!]
\includegraphics[scale=.45]{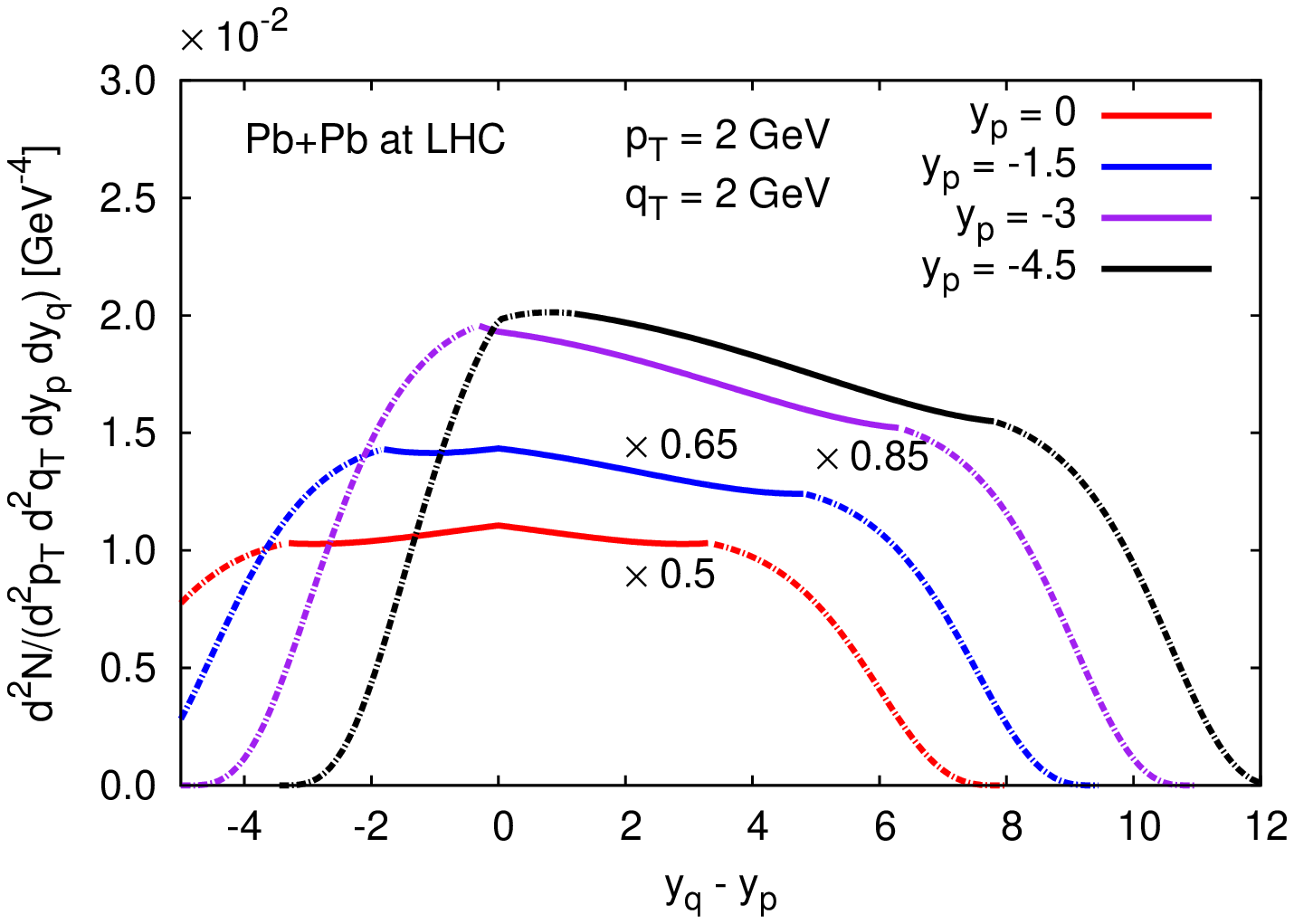}
\includegraphics[scale=.45]{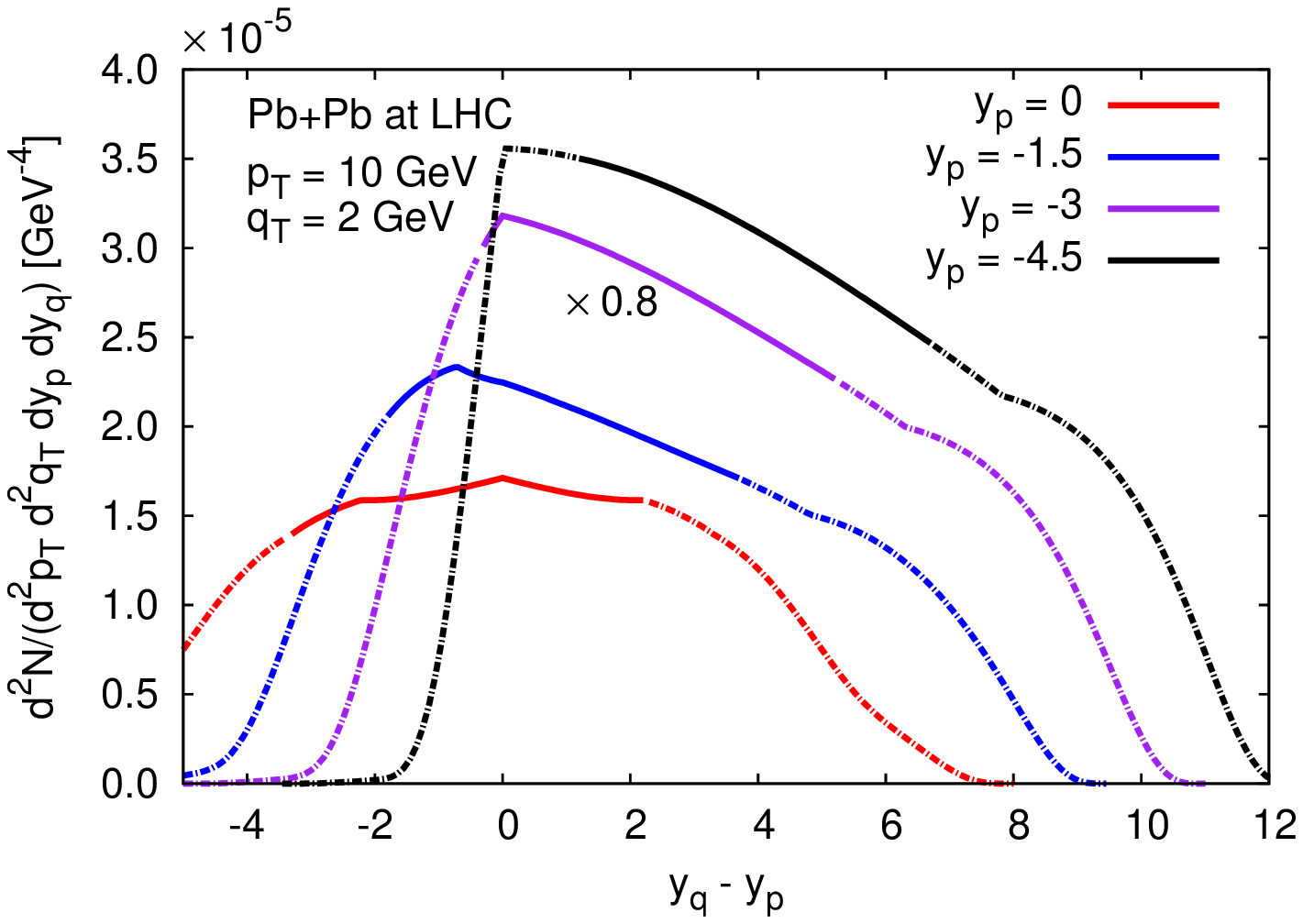}
\caption{\label{fig:lhc}The predicted two particle correlation spectrum as a function
  of the rapidity difference between the two gluons. The figure on the
  left corresponds to the case where the transverse momenta of the two
  gluons are equal and are $p_\perp,q_\perp =2$~GeV; the figure on the
  right depicts the case where $p_\perp=10$~GeV and $q_\perp = 2$~GeV.
  The different plots reflect different trigger rapidities. Solid
  parts of each curve correspond to $x<0.01$ in both nuclei; the
  dashed parts are sensitive to $x> 0.01$ in at least one of the
  nuclei.  We have rescaled some curves, by the given factors, for clarity.}
\end{figure}

At RHIC energies, the range in rapidity where the results are
sensitive to small $x$ physics exclusively is quite limited. At the
LHC, this range is much larger and the effects of QCD evolution on
long range rapidity correlations is more transparent. In
fig.~\ref{fig:lhc}, we show results for the two particle cumulant
$C({\bf p},{\bf q})$ as a function of the rapidity difference between
the two gluons. In the figure on the left, the correlation is plotted
for $p_\perp=q_\perp=2$~GeV; the right figure corresponds to the
asymmetric case of $p_\perp=10$~GeV and $q_\perp=2$~GeV.  For both
scenarios, we show the evolution in rapidity of the two particle
correlation at different trigger rapidities $y_p$.  The solid part of
each curve corresponds to the kinematic range where only $x\leq 0.01$
values in each of the nuclear wavefunctions are being probed. In
contrast, the dashed part of each curve denotes the kinematic range
which is sensitive to $x> 0.01$ for at least one of the nuclei; in
this regime, the results are more sensitive to the form chosen for the
large $x$ extrapolation than to the high energy QCD evolution equations at
small $x$.  Because of the large kinematic reach of the LHC, we
observe in fig.~\ref{fig:lhc} that we have a region contributing to the
double inclusive rapidity spectrum, of nearly $7$ units in the
rapidity difference of the two gluons, which is sensitive only to the
small $x$ evolution in the nuclear wavefunctions.  The shape and
magnitude of these correlations will therefore give us unique insight
into the evolution of multi-parton correlations in high energy QCD.

\section{Summary}
\label{sec:summary}

In Ref.~\cite{GelisLV5}, a general formula
(\eq{\ref{eq:double-inclusive-1}}) was derived for double inclusive
gluon production in the Glasma at arbitrary rapidity separations. In
this paper, we showed that this formula reduces to a compact
expression, \eq{\ref{eq:double-inclusive-4}}, in terms of the
unintegrated gluon distributions in the two nuclei. This
simplification holds when $p_\perp$, $q_\perp \gtrsim \qs$ and when the
mean field Balitsky-Kovchegov (BK) framework --- valid in the 
large $N_c$ limit --- is used to describe the high energy evolution
of the nuclei. The unintegrated gluon distributions at small $x$ are
simply related to the dipole foward amplitude, which in turn satisfies
the BK equation. Solving the running coupling form of the
BK equation, with initial conditions determined from fits to fixed
target e+A data, we computed the double inclusive spectrum at RHIC and
LHC energies. In the case of the former, we obtained a good agreement
with the PHOBOS data, albeit the kinematic region where small $x$
partons are probed in both nuclei is rather small. In the latter case,
we showed that there is a wide kinematic window for rapidity
correlations at the LHC. Our results therefore open a new window into
the study of the high energy evolution of multiparton correlations in nuclear
wavefunctions.

\section*{Acknowledgements}

We thank Javier Albacete, Adrian Dumitru, Anna Stasto and Kirill Tuchin for very
useful discussions. We are especially grateful to Gregory Soyez for
his co-ordinate space BK code. K.D. and R.V.'s research is supported
by the US Department of Energy under DOE Contract No.
DE-AC02-98CH10886. T.L. is supported by the Academy of Finland,
project 126604. F.G. is supported in part by Agence Nationale de la
Recherche via the program ANR-06-BLAN-0285-01.

\appendix

\section{Evaluation of \eq{\protect\ref{eq:F}}}
\label{sec:appendixA}
In this appendix, we work out some of the details of the derivation of
the double inclusive spectrum. In particular, \eq{\ref{eq:F}} is
expressed as the product of eight color charge densities. For a
non-local Gaussian distribution of these sources, one has nine
possible pairings of color source densities. These are evaluated
explicitly below.
\begin{eqnarray}
\mathcal{F}^{(9)}&=&(2\pi)^8 \delta^{bf}\delta^{dh}\delta^{cg}\delta^{ei}\delta(\kpn{1}-\kpn{2})\mu^2_{A_1}(y_p,\vert\kpn{1}\vert)\delta(\kpn{3}-\kpn{4})\mu^2_{A_1}(y_q,\vert\kpn{3}\vert)\nonumber\\
&\times&\delta(\kpn{2}-\kpn{1})\mu^2_{A_2}(y_p,\vert\pp-\kpn{1}\vert)\nonumber\\
&\times&\delta(\kpn{3}-\kpn{4})\mu^2_{A_2}(y_q,\vert\qp-\kpn{3}\vert)\; ,
\label{eq:F-9}
\end{eqnarray}
\begin{eqnarray}
\mathcal{F}^{(1)}&=&(2\pi)^8 \delta^{bf}\delta^{dh}\delta^{ce}\delta^{gi}\delta(\kpn{1}-\kpn{2})\mu^2_{A_1}(y_p,\vert\kpn{1}\vert)\delta(\kpn{3}-\kpn{4})\mu^2_{A_1}(y_q,\vert\kpn{3}\vert)\nonumber\\
&\times&\delta(\qp+\pp-\kpn{3}-\kpn{1})\mu^2_{A_2}(y_q,\vert\pp-\kpn{1}\vert)\nonumber\\
&\times&\delta(\qp+\pp-\kpn{4}-\kpn{2})\mu^2_{A_2}(y_q,\vert\pp-\kpn{2}\vert)\; ,
\label{eq:F-1}
\end{eqnarray}
\begin{eqnarray}
\mathcal{F}^{(2)}&=&(2\pi)^8 \delta^{bf}\delta^{dh}\delta^{ci}\delta^{ge}\delta(\kpn{1}-\kpn{2})\mu^2_{A_1}(y_p,\vert\kpn{1}\vert)\delta(\kpn{3}-\kpn{4})\mu^2_{A_1}(y_q,\vert\kpn{3}\vert)\nonumber\\
&\times&\delta(\pp-\qp-\kpn{2}+\kpn{3})\mu^2_{A_2}(y_q,\vert\pp-\kpn{2}\vert)\nonumber\\
&\times&\delta(\qp-\pp-\kpn{4}+\kpn{1})\mu^2_{A_2}(y_q,\vert\pp-\kpn{1}\vert)\; ,
\label{eq:F-2}
\end{eqnarray}
\begin{eqnarray}
\mathcal{F}^{(3)}&=&(2\pi)^8 \delta^{fh}\delta^{bd}\delta^{cg}\delta^{ie}\delta(\kpn{2}+\kpn{4})\mu^2_{A_1}(y_p,\vert\kpn{2}\vert)\delta(\kpn{1}+\kpn{3})\mu^2_{A_1}(y_p,\vert\kpn{1}\vert)\nonumber\\
&\times&\delta(\kpn{1}-\kpn{2})\mu^2_{A_2}(y_p,\vert\pp-\kpn{1}\vert)\nonumber\\
&\times&\delta(\kpn{3}-\kpn{4})\mu^2_{A_2}(y_q,\vert\qp-\kpn{3}\vert)\; ,
\label{eq:F-3}
\end{eqnarray}
\begin{eqnarray}
\mathcal{F}^{(6)}&=&(2\pi)^8 \delta^{bh}\delta^{df}\delta^{cg}\delta^{ie}\delta(\kpn{1}-\kpn{4})\mu^2_{A_1}(y_p,\vert\kpn{1}\vert)\delta(\kpn{2}-\kpn{3})\mu^2_{A_1}(y_p,\vert\kpn{2}\vert)\nonumber\\
&\times&\delta(\kpn{1}-\kpn{2})\mu^2_{A_2}(y_p,\vert\pp-\kpn{1}\vert)\nonumber\\
&\times&\delta(\kpn{3}-\kpn{4})\mu^2_{A_2}(y_q,\vert\qp-\kpn{3}\vert)\; ,
\label{eq:F-6}
\end{eqnarray}
\begin{eqnarray}
\mathcal{F}^{(5)}&=&(2\pi)^8 \delta^{fh}\delta^{bd}\delta^{ge}\delta^{ic}\delta(\kpn{2}+\kpn{4})\mu^2_{A_1}(y_p,\vert\kpn{2}\vert)\delta(\kpn{1}+\kpn{3})\mu^2_{A_1}(y_p,\vert\kpn{1}\vert)\nonumber\\
&\times&\delta(\pp-\qp-\kpn{2}+\kpn{3})\mu^2_{A_2}(y_q,\vert\pp-\kpn{2}\vert)\nonumber\\
&\times&\delta(\qp-\pp-\kpn{4}+\kpn{1})\mu^2_{A_2}(y_q,\vert\pp-\kpn{1}\vert)\; ,
\end{eqnarray}
\begin{eqnarray}
\mathcal{F}^{(7)}&=&(2\pi)^8 \delta^{fd}\delta^{bh}\delta^{ce}\delta^{gi}\delta(\kpn{2}-\kpn{3})\mu^2_{A_1}(y_p,\vert\kpn{2}\vert)\delta(\kpn{1}-\kpn{4})\mu^2_{A_1}(y_p,\vert\kpn{1}\vert)\nonumber\\
&\times&\delta(\qp+\pp-\kpn{3}-\kpn{1})\mu^2_{A_2}(y_q,\vert\pp-\kpn{1}\vert)\nonumber\\
&\times&\delta(\qp+\pp-\kpn{4}-\kpn{2})\mu^2_{A_2}(y_q,\vert\pp-\kpn{2}\vert)\; ,
\end{eqnarray}
\begin{eqnarray}
\mathcal{F}^{(4)}&=&(2\pi)^8 \delta^{fh}\delta^{bd}\delta^{ce}\delta^{gi}\delta(\kpn{2}+\kpn{4})\mu^2_{A_1}(y_p,\vert\kpn{2}\vert)\delta(\kpn{1}+\kpn{3})\mu^2_{A_1}(y_p,\vert\kpn{1}\vert)\nonumber\\
&\times&\delta(\qp+\pp-\kpn{3}-\kpn{1})\mu^2_{A_2}(y_q,\vert\pp-\kpn{1}\vert)\nonumber\\
&\times&\delta(\qp+\pp-\kpn{4}-\kpn{2})\mu^2_{A_2}(y_q,\vert\pp-\kpn{2}\vert)\; ,
\end{eqnarray}
and
\begin{eqnarray}
\mathcal{F}^{(8)}&=&(2\pi)^8 \delta^{fd}\delta^{bh}\delta^{ge}\delta^{ic}\delta(\kpn{2}-\kpn{3})\mu^2_{A_1}(y_p,\vert\kpn{2}\vert)\delta(\kpn{1}-\kpn{4})\mu^2_{A_1}(y_p,\vert\kpn{1}\vert)\nonumber\\
&\times&\delta(\pp-\qp-\kpn{2}+\kpn{3})\mu^2_{A_2}(y_q,\vert\pp-\kpn{2}\vert)\nonumber\\
&\times&\delta(\qp-\pp-\kpn{4}+\kpn{1})\mu^2_{A_2}(y_q,\vert\pp-\kpn{1}\vert)\; .
\end{eqnarray}

The classification of these contributions was examined previously in
\cite{DumitGMV1} in the framework of the MV model. The analysis is
identical here.  The expression $\mathcal{F}^{(9)}$ is trivial as it
cancels the square of the single particle distribution. Let us look at
the $\delta$-functions in $\mathcal{F}^{(4),(8)}$.  These yield a 
local $[\delta(\pp\pm\qp)]^2$-contribution that we shall neglect here
as in Ref.~\cite{DumitGMV1}.  Similarly,
expressions $\mathcal{F}^{(5),(7)}$ are sub-dominant\footnote{We note that there is an order one contribution coming from $\mathcal{F}^{(5),(7)}$ when the relative angle between $p_\perp, q_\perp$ is $\Delta\phi_{pq} \lesssim \frac{\qs}{p_\perp}$.  In the limit where $\frac{\qs}{p_\perp} \ll 1$ these contributions will be washed out by re-scattering in the same manner as the $\delta$-function contributions coming from $\mathcal{F}^{(4),(8)}$, and thereby not alter our result.  We thank Kirill Tuchin for pointing out this subtlety to us.}.  The leading
terms are therefore $\mathcal{F}^{(1),(2),(3),(6)}$. If we plug these
back into \eq{\ref{eq:double-inclusive-3}}, we obtain the following
four contributions to the two gluon spectrum
\begin{eqnarray}
C^{(1)}({\bf p},{\bf q})&=&
\frac{g^{12}N_c^2(N_c^2-1) S_\perp}{16}
\int \ud^2\kpn{1}\;\mathcal{G}({\bf p},{\bf q};\{\kpn{i}^{(1)}\})\nonumber\\
&\times&\mu_{A_1}^2(y_p,\vert\kpn{1}\vert)\mu^4_{A_2}(y_q,\vert\pp-\kpn{1}\vert)
\;\mu^2_{A_1}(y_q,\vert\pp+\qp-\kpn{1}\vert)\; ,
\label{eq:C-1}
\end{eqnarray}
where
$\{\kpn{i}^{(1)}\}\equiv\{\kpn{1},\kpn{1},\pp+\qp-\kpn{1},\pp+\qp-\kpn{1}\}$,
\begin{eqnarray}
C^{(3)}({\bf p},{\bf q})&=&
 \frac{g^{12}N_c^2(N_c^2-1) S_\perp}{16}
\int \ud^2\kpn{1}\;\mathcal{G}({\bf p},{\bf q};\{\kpn{i}^{(3)}\})\nonumber\\
&\times&\mu_{A_1}^4(y_p,\vert\kpn{1}\vert)\mu^2_{A_2}(y_p,\vert\pp-\kpn{1}\vert)
\;
\mu^2_{A_2}(y_q,\vert\qp+\kpn{1}\vert)\; ,
\label{eq:C-3}
\end{eqnarray}
where $\{\kpn{i}^{(3)}\}\equiv\{\kpn{1},\kpn{1},-\kpn{1},-\kpn{1}\}$,
\begin{eqnarray}
C^{(2)}({\bf p},{\bf q})&=&
\frac{g^{12}N_c^2(N_c^2-1) S_\perp}{16}
\int \ud^2\kpn{1}\;\mathcal{G}({\bf p},{\bf q};\{\kpn{i}^{(2)}\})\nonumber\\
&\times&\mu_{A_1}^2(y_p,\vert\kpn{1}\vert)\mu^4_{A_2}(y_q,\vert\pp-\kpn{1}\vert)
\;\mu^2_{A_1}(y_q,\vert\qp-\pp+\kpn{1}\vert)\; ,
\label{eq:C-2}
\end{eqnarray}
where
$\{\kpn{i}^{(2)}\}\equiv\{\kpn{1},\kpn{1},\qp-\pp+\kpn{1},\qp-\pp+\kpn{1}\}$,
and finally
\begin{eqnarray}
C^{(6)}({\bf p},{\bf q})&=&
\frac{g^{12}N_c^2(N_c^2-1) S_\perp}{16}
\int \ud^2\kpn{1}\mathcal{G}({\bf p},{\bf q};\{\kpn{i}^{(6)}\})\nonumber\\
&\times&\mu_{A_1}^4(y_p,\vert\kpn{1}\vert)\mu^2_{A_2}(y_p,\vert\pp-\kpn{1}\vert)
\;\mu^2_{A_2}(y_q,\vert\qp-\kpn{1}\vert)\; ,
\label{eq:C-6}
\end{eqnarray}
where $\{\kpn{i}^{(6)}\}=\{\kpn{1},\kpn{1},\kpn{1},\kpn{1}\}$.  Using
\eq{\ref{eq:unint-gluon-dist}}, we can express
$C^{(1)}+C^{(2)}+C^{(3)}+C^{(6)}$ as \eq{\ref{eq:double-inclusive-4}}.

\section{Initial conditions for BK evolution}

The initial conditions for BK evolution of protons and nuclei are
obtained by comparing results for the dipole cross-section to deep
inelastic scattering data. The inclusive structure function $F_2$ is
given by
\begin{eqnarray}
F_2^A(x,Q^2)=\frac{Q^2}{4\pi^2\alpha_{em}}\left(\sigma^T_{_A}+\sigma^L_{_A}\right)\, ,
\end{eqnarray}
where $\sigma^{T,L}_{_A}$ is the virtual photon-nucleus cross section for
transverse and longitudinal polarizations of the virtual photon. These in turn are given by
\begin{eqnarray}
\sigma^{T,L}_{_A}(x,Q^2)=\int_0^1 \ud z\int \ud^2{\bf b}\; \ud^2{\bf r}\; \vert\Psi_{T,L}(z,Q^2,{\bf r})\vert^2\mathcal{N}_{_A}({\bf b},{\bf r},x)\,,
\end{eqnarray}
where $\mathcal{N}_{_A}$ is the dipole-nucleus scattering amplitude.  We assume here that the ${\bf b}$ dependence can be factorized as 
\begin{eqnarray}
\mathcal{N}_{_A}({\bf b},{\bf r},x)=2\mathcal{T}_{_A}({\bf b})\,\mathcal{N}_{_A}({\bf r},x)
\end{eqnarray}

The virtual photon-nucleus cross section can then be expressed as 
\begin{eqnarray}
\sigma^{T,L}_{_A}(x,Q^2)=\sigma_{_A}\int_0^1 \ud z\int \ud^2{\bf r}\;\vert\Psi_{T,L}(z,Q^2,{\bf r})\vert^2\mathcal{N}_{_A}({\bf r},x)
\end{eqnarray}
where
\begin{eqnarray}
\sigma_{_A}=2\int \ud^2{\bf b}\; \mathcal{T}_{_A}({\bf b})
\end{eqnarray}

\subsection*{Initial condition for protons}

The initial condition for protons was determined from a global fit of
$F_2$ data in the work of \cite{AlbacAMS1}.  Two different models for
the initial condition were used in that work. The first is the GBW
model
\begin{eqnarray}
N(r,Y=0)=1-\exp\left[-\frac{Q^2_{s_0}r^2}{4}\right]
\end{eqnarray}
and the other is the MV model
\begin{eqnarray}
N(r,Y=0)=1-\exp\left[-\left(\frac{Q^2_{s_0}r^2}{4}\right)^{\gamma}\ln\left(\frac{1}{r^2\Lambda_{QCD}^2}+e\right)\right]
\end{eqnarray}
The fit parameters obtained in~\cite{AlbacAMS1} are summarized in the table
\ref{tab:albacete}.

\begin{table}[h]
\begin{center}
\begin{tabular}{l | c | c | c | c}
I.C. & $\sigma_p$ (fm$^2$) & $Q_{s0,p}^2$ (GeV$^2$) & $C^2$ & $\gamma$ \\
\hline
GBW & 3.159 & 0.24 & 5.3 & NA \\
MV & 3.277  & 0.15 & 6.5 & 1.13 \\
\end{tabular}
\end{center}
\caption{Parameters for the initial condition of the
 proton dipole cross section obtained in \protect\cite{AlbacAMS1}.}
\label{tab:albacete}
\end{table}

\subsection*{Initial condition for nuclei}

We shall now consider the initial conditions for nuclei using the same
model as the initial conditions for protons.  We do not attempt to
perform a global fit since the data for DIS off nuclei are not nearly
of the same quality.  We use a model where the initial saturation
scale scales linearly with $A^{1/3}$,
\begin{eqnarray}
Q^2_{s_0}=c\,A^{1/3}Q^2_{s_0,p}\,,
\label{eq:satA}
\end{eqnarray}
where $c$ is a constant to be determined from the data.

\begin{figure}
\includegraphics[scale=.55]{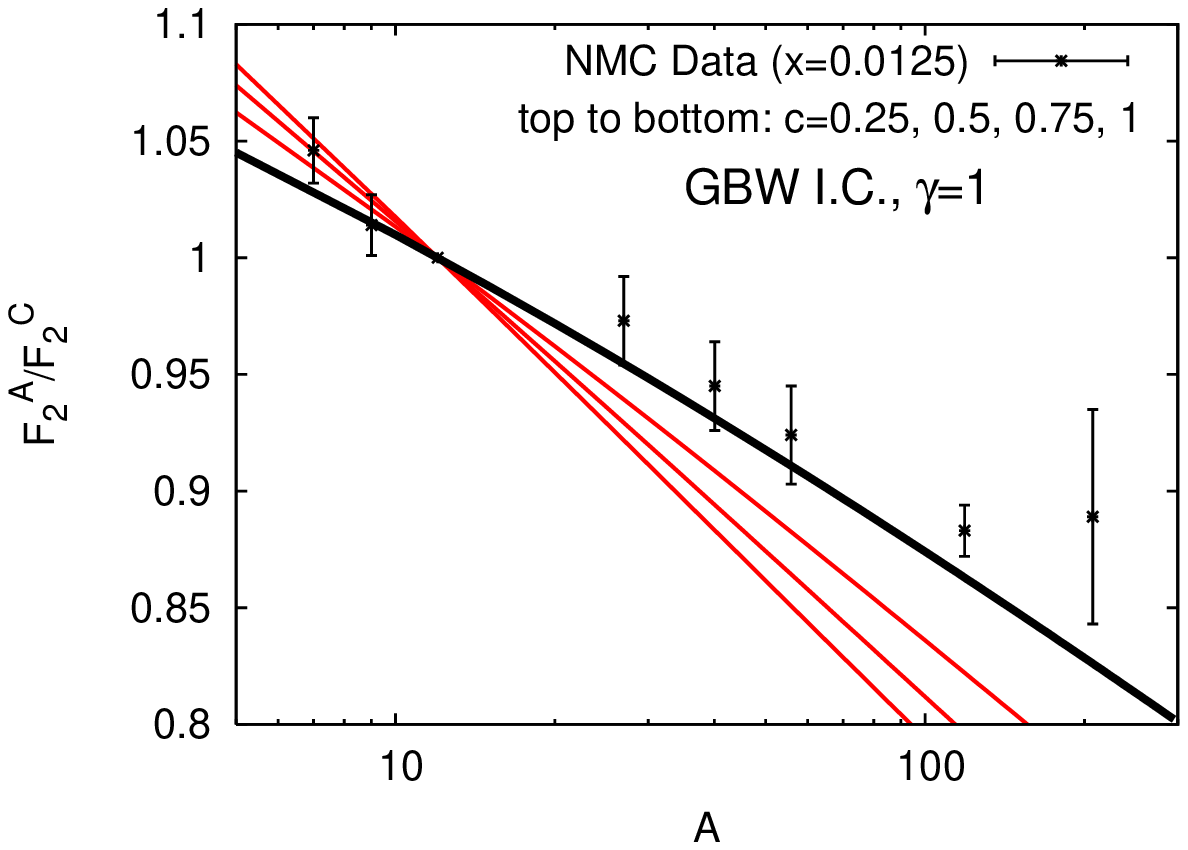}
\includegraphics[scale=.55]{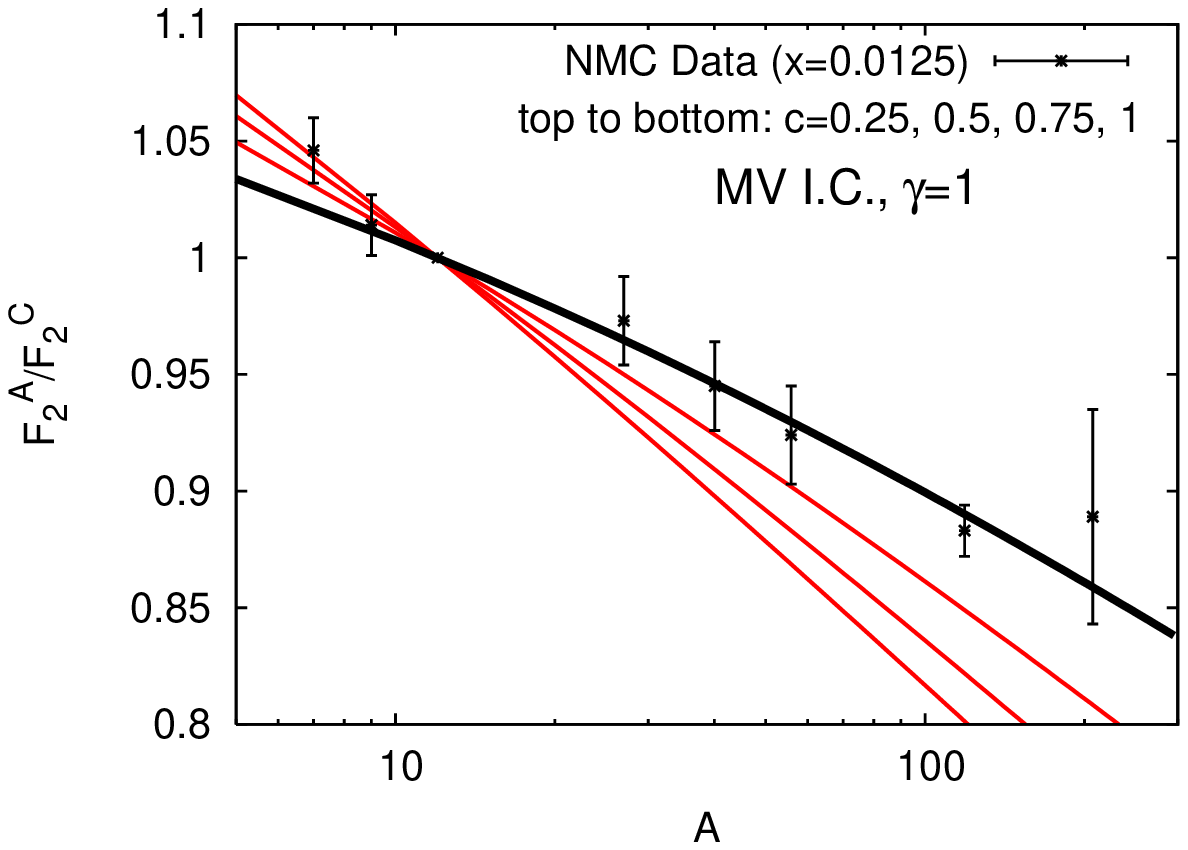}
\caption{DIS fixed target e+A data on the ratio of structure functions
  as a function of $A$ for fixed $x=0.0125$. The plots correspond to
  (right) MV model initial condition having $\gamma=1$ and (left) GBW
  initial condition.}
\label{fig:NMCAg1GBW}
\end{figure}

In order to constrain the initial condition, we begin by looking at
the New Muon Collaboration's (NMC) data~\cite{ArneoA1} for
$F_2^A/F_2^C$ as a function of $A$ at $x=0.0125$ which is close to
our $x_0=0.01$.  In this case, there is no BK evolution, and we have a
direct comparison of the nuclei's initial condition with the data.  In
computing $F_2^A/F_2^C$ we will need a model for how the cross section
scales with $A$.  We take $\sigma_{_A} = \left(\frac{A}{12}\right)^{2/3}
\sigma_{_C}$.  Figure \ref{fig:NMCAg1GBW} shows the NMC data as a
function of $A$ for the GBW initial condition for four different
values of $c$ (left) and the MV model initial condition having
anomalous dimension $\gamma=1$ (right).  It is clear that in order to
be consistent with the data we must take $c\approx 0.25$. The nuclear
saturation scale given by \eq{\ref{eq:satA}} is too small to be
consistent with measurements by other groups.

In \fig\ref{fig:NMCAg}, we show the NMC data on the ratio of structure
functions as a function of $A$ now using the MV initial condition with
anomalous dimension $\gamma=1.13$.  We find that for $c=0.5$ this 
fits the data rather well.  Based on the above results, for nuclei
we will use the MV model and take $\gamma=1.13$ and $Q^2_{s_0}=0.5
A^{1/3}Q^2_{s_0,p}$.  Therefore we have $Q^2_{s_0}=0.17,\ 0.26,\ 0.37$ and
$0.44$ (GeV)$^2$ for C, Ca, Sn and Au respectively.  Note that these
values of the saturation scale are for quarks in the fundamental
representation.  For gluons in the adjoint representation, the
corresponding saturation scale is
\begin{eqnarray}
\left( {\qs}^2 \right)_g=\frac{N_c}{C_{_F}}\left({\qs}^2\right)_q=2.25\left({\qs}^2\right)_q
\end{eqnarray}

\begin{figure}
\includegraphics[scale=.55]{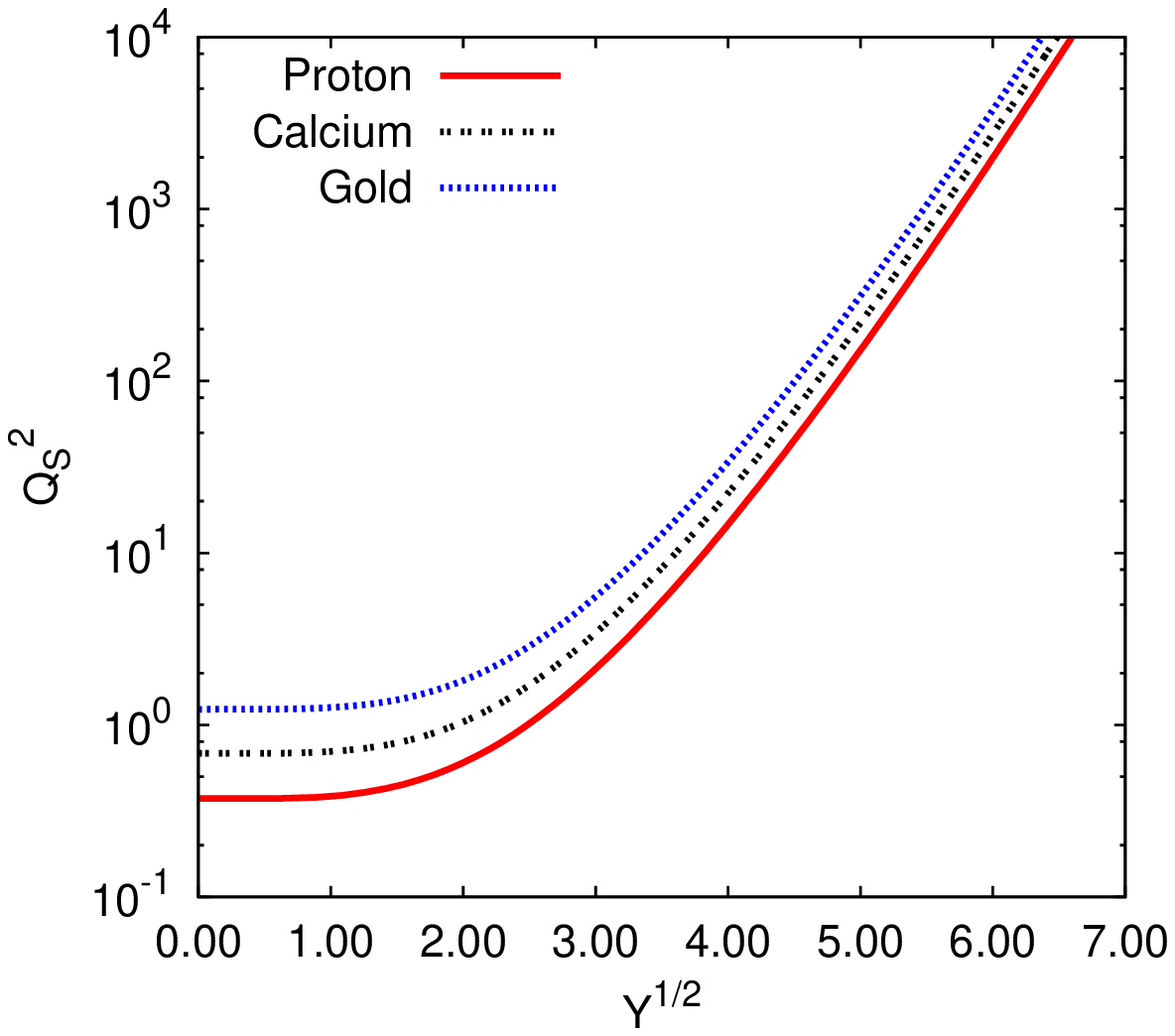}
\includegraphics[scale=.55]{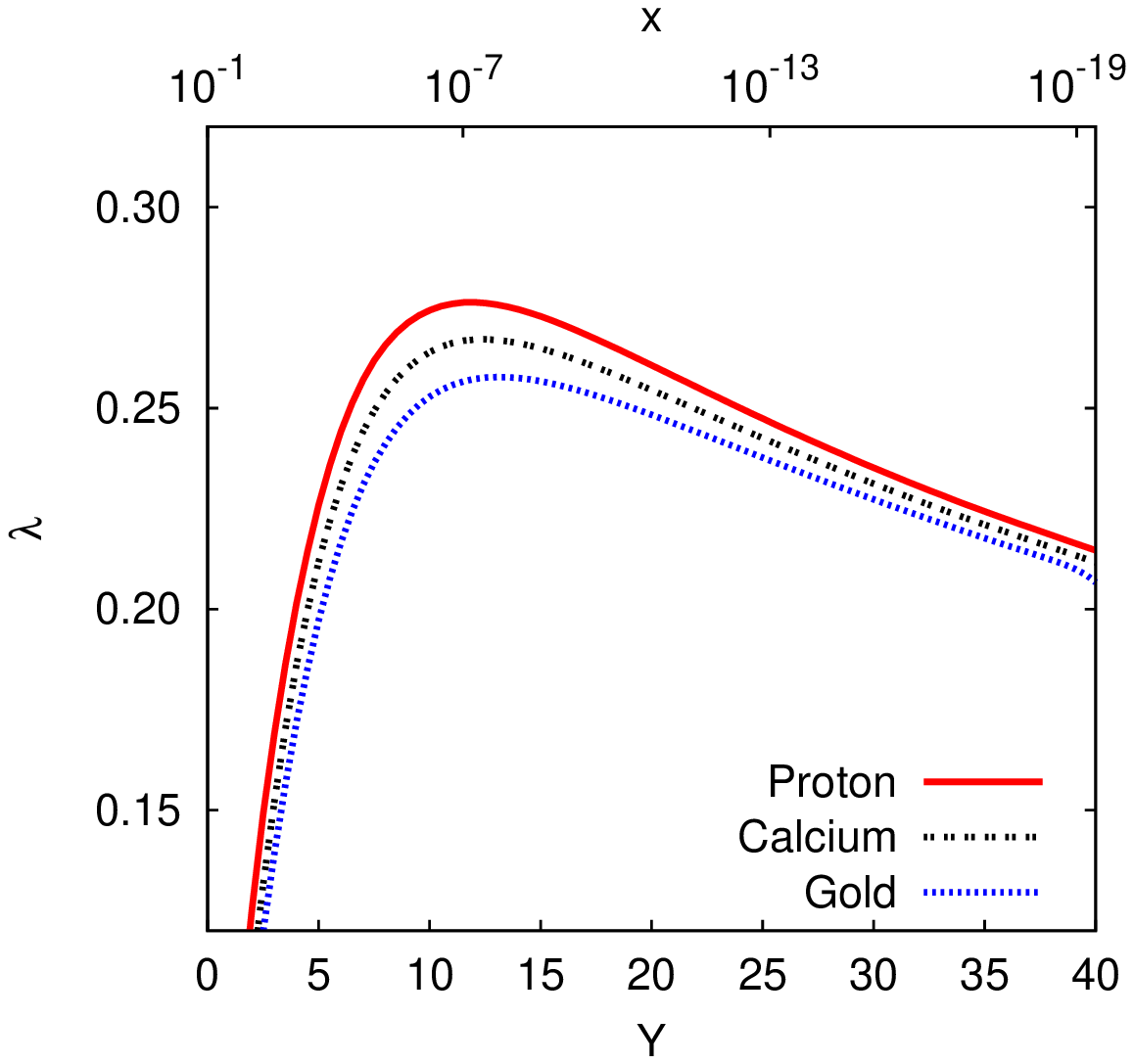}
\caption{Left: The saturation scale $\qs$ as a function of the square
  root of the rapidity $Y^{1/2}$ for protons, calcium and gold nuclei.
  Note that the slopes of the three curves approach the same value at
  large $Y$. Right: $\lambda = \ud\ln\qs^2/\ud Y$ as a function of $Y$
  approaches a universal value at large $Y$.}
\label{fig:QS}
\end{figure}

For gold nuclei this yields $({\qs}^2)_g\approx 1$~GeV$^2$ at $x=0.01$,
in fairly close agreement to the value of $1.3$~GeV$^2$ obtained in
\cite{KowalLV1,KowalT1}. Finally, we plot in \fig\ref{fig:QS} (left)
the saturation scale in the running
coupling case as a function of $Y^{1/2}$ for the proton, calcium and
gold nuclei. The behavior at small $Y^{1/2}$ is sensitive to the
initial conditions of each of these nuclei; however, at large
$Y^{1/2}$ (small $x$) the curves of the three nuclear approach the
same slope, as one expects asymptotically for the behavior of $\qs$
when running coupling effects are accounted for (see also
\cite{AlbacAMSW1}).  The same trend can be observed by plotting
(see \fig\ref{fig:QS} right)
$\lambda = \ud\ln \qs^2/\ud Y$, the parameter that sets the rate at which the
dipole amplitudes evolve with rapidity.  These results confirm the
universal behavior at large $Y$ predicted in Ref.~\cite{muell7}.

\end{document}